%% file: main.tex
\documentclass[conference]{IEEEtran}
\usepackage[english]{babel}
\usepackage[utf8x]{inputenc}
\usepackage[T1]{fontenc}
\usepackage{tikz}
\usepackage{graphicx}
\usetikzlibrary{positioning}
\usepackage{todonotes}
\usepackage{hyperref}
\usepackage{listings}
\usepackage{xspace}
\usepackage{ifthen}
\usepackage{comment}
\usepackage{amsmath}
\usepackage{cleveref}
\usepackage{ulem}
\usepackage{amsmath}
\usepackage{amsfonts}
\usepackage{subcaption}
\renewcommand{\emph}{\textit}
\renewcommand{\paragraph}[1]{\noindent\textbf{#1}}
\usepackage{amsthm}
 
\theoremstyle{definition}
\newtheorem{definition}{Definition}[section]

\usepackage[T1]{fontenc}

\graphicspath{{./Figures/}}

\usepackage{acro}
\acsetup{first-long-format=\itshape}
\include{acronyms}

\newcommand{\says}[3]{}

\newcommand{\luis}[1]{\says{Luis}{cyan}{#1}}
\newcommand{\matt}[1]{\says{Matt}{green}{#1}}

\begin{document}
\IEEEoverridecommandlockouts
    \title{Let's Talk Through Physics! Covert Cyber-Physical Data Exfiltration on Air-Gapped Edge Devices }
    
\author{
\IEEEauthorblockN{Matthew Chan}
\IEEEauthorblockA{\textit{Rutgers University}\\
matthew.chan@rutgers.edu}
\and
\IEEEauthorblockN{Nathaniel Snyder}
\IEEEauthorblockA{
\textit{University of California, Los Angeles}\\
natsnyder1@g.ucla.edu}
\and
\IEEEauthorblockN{Marcus Lucas}
\IEEEauthorblockA{
\textit{University of California, Los Angeles}\\
maluc@g.ucla.edu}
\and
\IEEEauthorblockN{Luis Garcia}
\IEEEauthorblockA{
\textit{University of California, Los Angeles}\\
lgarcia@isi.edu}
\and
\IEEEauthorblockN{Oleg Sokolsky}
\IEEEauthorblockA{
\textit{University of Pennsylvania}\\
sokolsky@cis.upenn.edu}
\and
\IEEEauthorblockN{James Weimer}
\IEEEauthorblockA{
\textit{Vanderbilt University}\\
weimerj@seas.upenn.edu}
\and
\IEEEauthorblockN{Insup Lee}
\IEEEauthorblockA{
\textit{University of Pennsylvania}\\
lee@cis.upenn.edu}
\and
\IEEEauthorblockN{Paulo Tabuada}
\IEEEauthorblockA{
\textit{University of California, Los Angeles}\\
tabuada@ee.ucla.edu}
\and
\IEEEauthorblockN{Saman Zonouz}
\IEEEauthorblockA{
\textit{Georgia Institute of Technology}\\
saman.zonouz@gatech.edu}
\and 
\IEEEauthorblockN{Mani Srivastava}
\IEEEauthorblockA{
\textit{University of California, Los Angeles}\\
mbs@ucla.edu}
}

    \maketitle
    \input{0-abstract}
    \input{1-introduction}
    \input{2-background}

    \input{3-overview}
    \input{4-design}
    \input{5-evaluation}
    \input{6-related-work}
    \input{7-discussion.tex}

    \input{8-conclusion}

    \bibliographystyle{acm}
    \bibliography{references}

\end{document}

%% file: 0-abstract.tex
\begin{abstract}
\luis{We need to change the story here to be more about M2M vulnerabilities, i.e., machines performing state estimation on each other for accountability.}
Although organizations are continuously making concerted efforts to harden their systems against network attacks by \textit{air-gapping} critical systems, attackers continuously adapt and uncover \textit{covert channels} to exfiltrate data from air-gapped systems. For instance, attackers have demonstrated the feasibility of exfiltrating data from a computer sitting in a Faraday cage by exfiltrating data using magnetic fields. Although a large body of work has recently emerged highlighting various physical covert channels, these attacks have mostly targeted open-loop cyber-physical systems where the covert channels exist on physical channels that are not being monitored by the victim. Network architectures such as fog computing push sensitive data to cyber-physical edge devices--whose physical side channels are typically monitored via state estimation. In this paper, we formalize covert data exfiltration that uses existing cyber-physical models and infrastructure of individual devices to exfiltrate data in a stealthy manner, i.e., we propose a method to circumvent cyber-physical state estimation intrusion detection techniques while exfiltrating sensitive data from the network.  


We propose a generalized model for encoding and decoding sensitive data within cyber-physical control loops. 
We evaluate our approach on a distributed IoT network that includes computation nodes residing on physical drones as well as on an industrial control system for the control of a robotic arm. Unlike prior works, we formalize the constraints of covert cyber-physical channel exfiltration in the presence of a defender performing state estimation.

\end{abstract}

%% file: 1-introduction.tex
\section{Introduction}

\luis{Again, right now the focus is a bit too much on air-gapping. We can following the following storyline to convert the message: state-estimation is critical in cyber-physical M2M networks to hold each device accountable (cite a bunch of autonomous IDS solutions.); traditional cyber-security techniques air-gapping techniques are not sufficient to secure information leakage (cite all the out-of-band attacks below). }
The practice of air-gapping critical systems--i.e., physically isolating a computer from an unsecured network--provides assurances against standard network vulnerabilties and significantly reduces the associated attack surface.  These defenses typically only break down in the face of insider or physical attacks as in the Stuxnet malware~\cite{falliere2011w32}. However, recent attacks have overcome air-gap defenses to exfiltrate data via covert channels that exploit device peripherals, including electromagnetic emanations~\cite{guri2014airhopper,guri2015gsmem,guri2016usbee}, magnetic fields~\cite{guri2018odini,guri2018magneto}, power consumption~\cite{guri2018powerhammer}, acoustic noise ~\cite{guri2017acoustic,guri2016fansmitter,song2016my,guri2018mosquito}, observable characteristics~\cite{gur2018xled,guri2017led,guri2019air}, and thermal emissions~\cite{guri2015bitwhisper}. The countermeasures proposed for such attacks typically discuss procedural countermeasures such as secure practices in the work environment or technological approaches that attempt to conceal or shield the physical covert channels. Mitigating the physical covert channels is feasible for static scenarios, e.g., data centers for a distributed cloud computing architecture. However, distributed computation architectures such as fog computation have evolved to perform computation on much more dynamic and adaptive edge network devices. Edge devices in the wild that are sensing and actuating in the physical environment with distributed state estimation introduce physical covert channels that exhibit much more complexities than the aforementioned channels.

\luis{I believe the edge computation story can be part of the M2M state estimation story, i.e., the distributed nature of IoT networks has called for a push to compute on the edge for reasons XYZ. The computation on the edge enables M2M state estimation...}
There have been several driving factors that have pushed the cloud computation paradigm to distributed computing on the \emph{edge}. In particular, the explosion of the internet of things (IoT) has called for an increased emphasis on the collateral attributes of distributed edge networks, e.g., mobility, wide-spread geographical location, low-latency, and heterogeneity~\cite{bonomi2012fog}. In parallel, recent works~\cite{yao2018fastdeepiot,esmaeilzadeh2012neural} have shown the feasibility of driving computation to edge devices in an attempt to facilitate advancements that target these attributes. Yet as applications are driven further from cloud computing towards the edge, there exists a tradeoff in utility versus physical security guarantees. In particular, emerging scenarios that rely on deployable and/or mobile infrastructures such as emergency response necessitate a means of distributed edge computation on devices that are cyber-physically insecure. As opposed to cloud data centers, these low-level devices are physically exposed and may not be able to deploy any of the aforementioned procedural countermeasures while providing cyber-physical runtime guarantees. However, unlike the aforementioned covert channels, these cyber-physical systems are typically monitored via supervisory controller state estimation to ensure the system is behaving correctly.  Therefore, an attacker's encoding mechanism for data exfiltration would need to be designed so as not to have the state estimator raise any flags. Because fog architectures are running inferencing closer to or on the edge devices, an attacker may have access to higher-level information inferred from the data and, as such, has to encode less bits into an attack since more information can be encoded into each bit. For instance, a drone that is monitoring a group of soldiers may have an inference algorithm that detects how many soldiers are in its view. An attacker would only have to encode the number of soldiers into the data exfiltration as opposed to sending the raw data.

In this paper, we show how an edge device's physical actuation can be used as a covert channel to exfiltrate sensitive data. In particular, we introduce a cyber-physical encoding technique that maintains \textit{stealthiness} against an entity who is monitoring the cyber-physical system via state estimation techniques. We begin by characterizing control system models for both an attacker and a defender in the same cyber-physical context.  We empirically demonstrate how an attacker would maximize the rate of transmission while maintaining stealthiness with respect to the physical covert channel. This also implies that our approach maintains the utility of the cyber-physical application. For instance, to encode data into the actuation of a drone, our approach would encode data into the movement of the drone while ensuring that the drone completes its waypoint navigation correctly.  This approach is analogous to prior attacks that focused on the semantic models of autonomous systems, e.g., cyber-physical attacks that target state-estimation techniques or adversarial machine learning techniques that target learned models. 

We evaluate our attack on two exemplary cyber-physical systems: a robotic arm in the context of an industrial control system as well as a drone surveilling an area of interest. For each system, we encode the data across a variety of applications and evaluate the efficacy of each attack. We use computer vision techniques to observe the physical actuation and decode the encoded bits. We also evaluate each attack against defenders with varying levels of probabilistic certainty about the estimated system states, including a ``perfect" defender that has access to the precise attacker model.  We optimize our attacks against state-of-the-art state estimation techniques and show how we would maximize transmission rate for each case with respect to the state estimation noise. We further enumerate countermeasures that can be embedded into state estimation techniques as well as the associated control mechanisms. 

Our contributions are summarized as follows:
\begin{itemize}
    \item We characterize the state-of-the-art of cyber-physical data exfilration techniques (Section~\ref{sec:background} and introduce a generalized cyber-physical covert channel attack model for data exfiltration that is optimized against state estimation techniques to maximize the transmission rate while maintaining stealthiness (Sections~\ref{sec:overview} and ~\ref{sec:design}).
    \item We evaluate our approach on a variety of applications across two exemplary cyber-physical systems and show the efficacy of such an attack (Section~\ref{sec:eval}).
    \item We discuss future directions of such attacks and enumerate countermeasures (Sections ~\ref{sec:related} and ~\ref{sec:conclusion}).
\end{itemize}

The source code and datasets of our system are available online at: \texttt{[repository]}\footnote{The source code and datasets will be available after the paper is published to respect the double-blind policy.}

%% file: 2-background.tex
\section{Background}\label{sec:background}
In this section we provide a background on air-gapped covert data exfiltration. We then discuss the recent advancements in edge computation that have driven sensitive data and applications to the edge.  to establish a preliminary foundation for generalizing a system model and its associated threat model.

\luis{The first subsection shoudl be about CPS vulnerabilities in M2M. We can then convert the Air-gapped covert data exfiltration to "out-of-band" vulnerabilties; We can probably move the current "computation on autonomous edge devices" subsection to the front and chagne the story based on the same chagnes I mentioned in the intro.}
\subsection{Air-Gapped Covert Data Exfiltration}
\luis{Change this to out-of-band data exfiltration.}
Currently, covert data exfiltration works have shown how physical side channels may be enabled across different modalities for air-gapped systems such as electromagnetic radiation~\cite{guri2014airhopper, guri2015gsmem,guri2016usbee}, magnetic fields~\cite{guri2018odini,guri2018magneto}, power consumption~\cite{guri2018powerhammer}, acoustic channel~\cite{faruque2016acoustic, guri2017acoustic,guri2018mosquito}, optical field~\cite{gur2018xled,guri2017led,guri2019air}, as well as thermal emissions~\cite{guri2015bitwhisper}. In all cases, these systems typically propose a cyber-physical air-gapped covert channel followed by an associated countermeasure to prevent such channels from being exploited. Subsequent works will then continue this attacker-defender game where a new covert channel is proposed to attack the hardened system. For instance, to provide a defense against the aforementioned attacks where data was exfiltrated via electromagnetic radiation~\cite{guri2014airhopper, guri2015gsmem,guri2016usbee}, technical countermeasures are proposed such as physical insulation and software-based reductions of information-bearing emissions. Subsequent attacks then proposed a means of circumventing the physical insulation of electromagnetic radiation by exfiltrating via the magnetic field emissions of the targeted device~\cite{guri2018odini,guri2018magneto}. 

The procedural and technical countermeasures presented the aforementioned attacks generally propose insulation of the physical channels in which data can be exfiltrated that are subsequently exploited. For both attacks and defenses, these approaches fail to encapsulate the physical model of these channels that stem from the memory-mapped inputs and outputs of the system. Such physical models can be used to perform cyber-physical state estimation to understand what will be the physical impact of a particular action in the cyber space. Further, state estimation allows for providing an understanding of the mutual dependency between physical channels, e.g., the correlation between a computer's fan operation and the acoustic channel. From a defender's perspective, state estimation not only enables the cyber-physical noise models that may need to be insulated, but also can perform intrusion detection if an attacker is explicitly encoding data into a particular channel that deviates from the estimated state of the channel. From an attacker's perspective, state estimation techniques can be used to craft complex cyber-physical attacks on neglected physical channels. In both cases, the respective problems are exacerbated when moving from the static, immobile systems considered in these works--e.g., data center computers that are easier to physically insulate--to mobile and autonomous edge devices that are difficult to physically insulate and expose even more cyber-physical channels. In this paper, we aim to formalize the notion of securing \textit{all} physical covert channels, particularly in the context of mobile and autonomous edge devices. 

  
\subsection{Computation on Autonomous Edge Devices}
Although edge and fog computation can be alluded to interchangeably~\cite{shi2016edge}, we refer to edge computation as the enabling technologies that perform data processing on devices that reside a single ``hop" away from sensors and actuators, i.e., directly interfacing with sensors and actuators. This implies that the edge devices will need to perform local processing of data in addition to maintaining any cyber-physical functions. The need for such edge computation stems from several factors, including the bottleneck and insecurity of networking, the inefficiency of cloud computation for real-time systems, as well as the fact that edge devices now \emph{produce} data instead of just consuming data~\cite{shi2016edge}. 

With the increasing demand for such frameworks, the industry has been quick to provide IoT edge services that excel in different domains. Platforms such as Microsoft Azure IoT Edge~\cite{gremban2019}, AWS IoT Greegrass~\cite{kurniawan2018learning}, and Watson IoT~\cite{watsonIoT} have enabled previous cloud services to be migrated to the edge devices in collaboration with cloud services. Google IoT Edge~\cite{google} has similarly enabled and facilitated machine learning on the edge. GE Predix~\cite{gepredix} has enabled distributed edge services for Industrial IoT (IIoT) applications. The increasing ubiquity of such technologies has spilled into privacy-sensitive edge applications that call for increased security and privacy measures.

\begin{figure*}[htp!]
  \centering
  \includegraphics[width=0.8\textwidth]{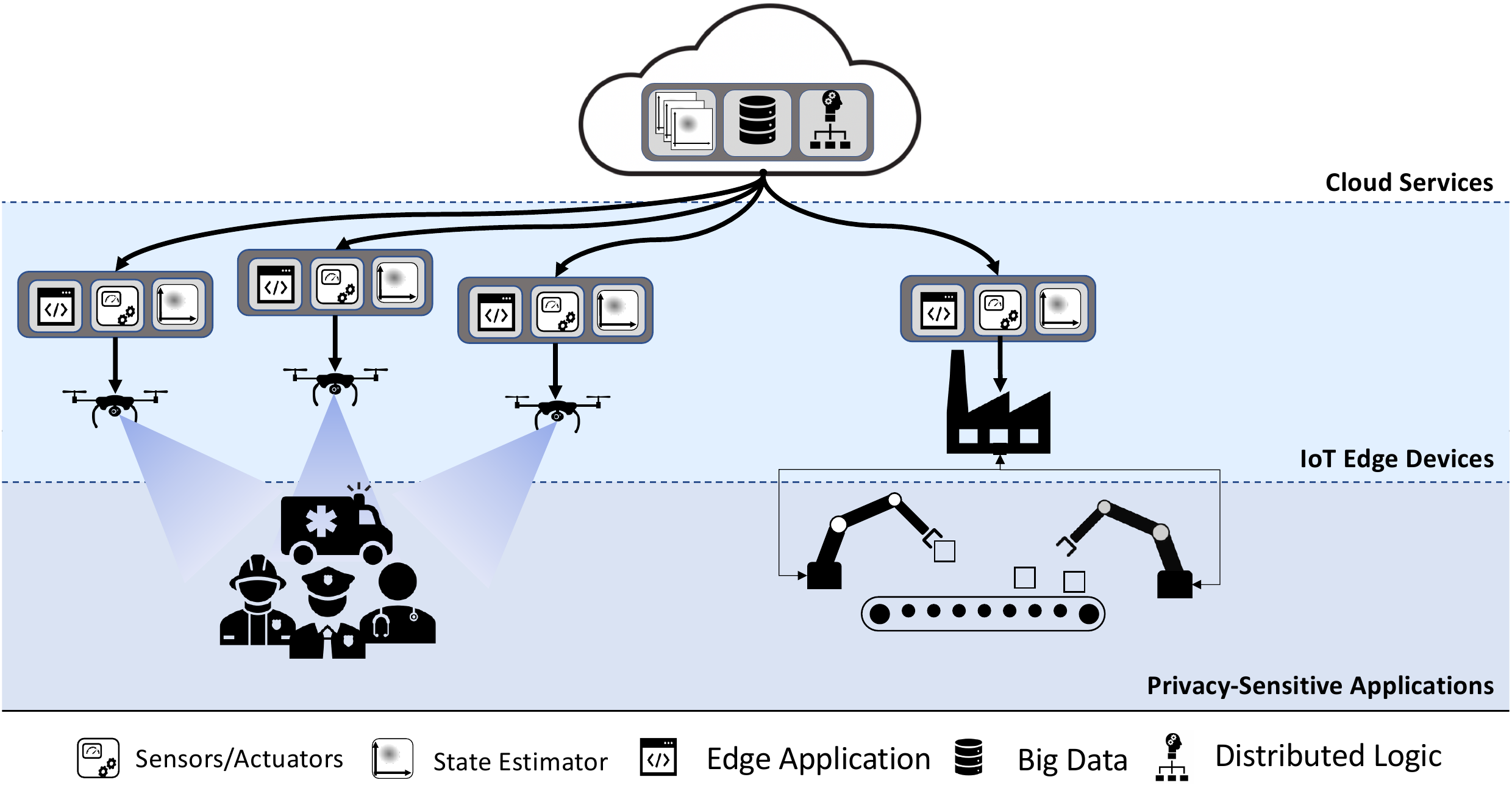}
   \caption{Edge computation system model.\luis{This figure should be changed to reflect the cyber-physical M2M system model.}  } 
  \label{fig:system-model}
  \vspace{-.05in}
\end{figure*} 
\noindent\textbf{Privacy-sensitive edge applications.} Prior works have shown that computation on the edge can significantly reduce the latency for invasive applications such as facial recognition~\cite{yi2015fog} or cognitive assistance~\cite{ha2014towards}. Such local processing reduces security and privacy concerns for sensitive contexts such as emergency response scenarios where IoT devices may provide supportive services for humans~\cite{srinivasan2016privacy}, as shown in Figure~\ref{fig:system-model}. However, enabling such inference abstractions on the edge now exposes the higher level logic that can be inferred from the raw data that is being processed. In distributed cloud computation models, the raw data (e.g., an image) was sent over the network to be processed on the cloud. Exfiltrating raw data through physical side channels is more challenging as each bit represents a very small fraction of the larger signal. However,  raised abstractions and inference applications enable more information to be encoded into each bit that is exfiltrated.
We now characterize the models and assumptions for physical covert channels from both an attacker's perspective as well as a defender of the system.

%% file: 3-overview.tex
\section{Models and Assumptions}\label{sec:overview}
In this section we will provide a precise system model considered in this paper for physical covert channels. We then define the threat model along with the adversarial assumptions. In particular, we categorize the different attack scenarios that arise in this context.

\begin{figure}[htp]
  \centering
@  \includegraphics[width=\linewidth]{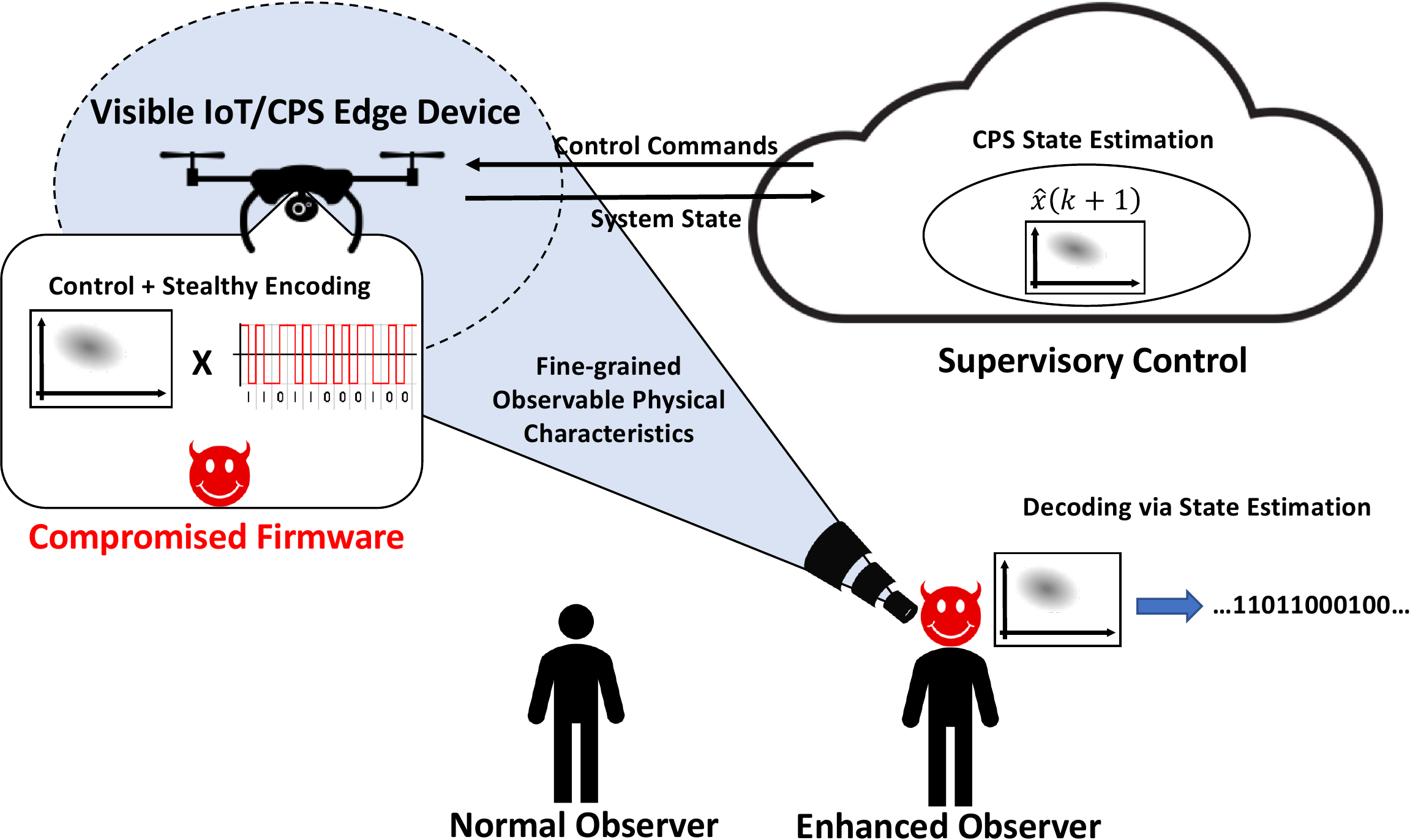}
   \caption{Data exfiltration attack overview.\luis{This figure should be updated to replace the humans and the supervisory control with other autonomous devices. Depending on the system model figure, we can have a network of CPS M2M and show how data can flow from non-montiroed variables.}} 
  \label{fig:attack-model}
  \vspace{-.05in}
\end{figure} 
\subsection{System Model}
The system model we consider in this paper is depicted in Figure \ref{fig:attack-model}, where the visible IoT/CPS Edge device is monitoring a sensitive application in the context of an edge computation system model as depicted in Figure \ref{fig:system-model}. A supervisory controller is monitoring the system state of the edge device and sending high-level control commands accordingly, e.g., an air traffic controller sending a coordinate setpoint for a drone. The supervisory controller is also using the state information to ensure that the system state is consistent with previously sent control commands, e.g., a drone's previous state has been updated according to the physical dynamics. We assume that the edge device has local control loops that convert the high-level commands from the supervisory controller to local actuation with respect to its internally maintained state estimation, e.g., a drone's stability and waypoint navigation control loops. Finally, we assume that there may be one or more humans in the same vicinity that can observe the physical characteristics of the device from a distance. The notion of a human's perception is an analog to a human's perception with respect to distortion models in the context of adversarial machine learning~\cite{papernot2016limitations}. We now discuss the threat model with respect to this system model.
\subsection{Threat model} 
The threat model has two components: the compromised edge device that is encoding sensitive information into physical actuation, and an adversarial observer that is decoding the encoded actuation.

\noindent\textbf{Compromised software model.} We assume that an attacker has compromised the edge device in such a way in which the attacker has full control of one or more physical actuators of the device. We also assume that the attacker has access to the physical dynamics of the edge device such that an attacker can model the state estimator along with an estimated noise model for the device. Such assumptions have been in used in prior cyber-physical state estimation attacks as these models can be practically obtained~\cite{garcia2017hey}. However, unlike previous cyber-physical system attacks, we do not assume that the attacker can report false system states to the supervisory controller as the mechanism that reports the sensed system state may be located on a different chip than the exploited software module, e.g., an attacker who has compromised the GPIO microcontroller may not be able to compromise the reported values of a separate GPS chip that is reporting the location.

\noindent\textbf{Enhanced adversarial observer model.} For our enhanced adversarial observer, we simply assume that an adversary may be able to ``zoom" in on the edge device to observe fine-grained observable physical characteristics that would not be observable to a normal human observer. 
To formalize the adversary model, we place both the attacker and the defender in a control systems context.

\subsection{A Control Systems Summarization}
We formalize a control-theoretic, systems-oriented model of our proposed attacker and defender models. This systems-view summarizes the aforementioned attack vectors and elaborates on how an attacker/defender would begin to model the physical variables and their cyber-physical dependencies. To start, we can choose to describe the cyber-physical system as a set of discrete-time, non-linear stochastic equations representing the dynamics of the system as:

\begin{equation}
    x(t+1) = f(x(t),u(t)) + g(w(t))
\end{equation}
With $x(t)\in \mathbb{R}^n$ as the state vector, $u(t)\in R^p$ as the control input, $f(x(t),u(t))$ as a deterministic propagation function and $g(w(t))$ being a potentially non-linear function of the system's process noise $w(t)\in R^r$, described by an underlying probability density function~\cite{gustafson1975design}. 

The CPS can also access information from its available set of sensing instruments. We can model a set of sensors using a stochastic transformation over the system state vector as:
\begin{equation}
    z(t) = h(x(t)) + v(t)
\end{equation}

With $z(t)\in \mathbb{R}^q$ as the sensor measurement vector received from the sensing instruments and $v(t)\in R^q$ representing a stochastic noise term, commonly regarded to be independent of the process noise $w(t)$.

Such a set of non-linear system equations can be linearized about a known system equilibrium-point (or the system's current state $\hat{x}(t)$) as: $\delta x(t)$ =  $x(t) - x_{eq}(t)$, with $\delta x(t) \in \mathbb{R}^{n}$ and expressed linearly as:

\begin{equation}
    \delta x(t+1) = A(t)\delta x(t) + B(t)\delta u(t) + G(t)w(t)
\end{equation}
With $A(t), B(t), G(t)$ possibly time varying (or time-invariant). We consider $A \in R^{n\times n}$ to represent the state dynamics matrix, $B \in \mathbb{R}^{n\times p}$, as the control input matrix, and $G \in \mathbb{R}^{n\times r}$ as a process-noise propagation matrix, respectively.

Using the dynamics models and on board measurement equipment, we can develop control structures and state estimators, possibly linear or non-linear in nature, to describe the attack or defense strategy of an adversary or defender. In constraining the system dynamics to a set of linearized propagation equations, when necessary, well-known estimators and controllers such as Kalman Filters and Linear Regulators can be applied to model simple and descriptive representations of the respective adversary and defender strategies in consideration. By subjecting the defender and attacker models to a control-theoretic perspective, we can provide provable measures, when necessary, over the various ``blocks" of the system, i.e., the adversary, the defender, and the model dynamics which is referred to as the \text{plant}.

\begin{figure}[tp]
  \centering
  \includegraphics[width=\linewidth]{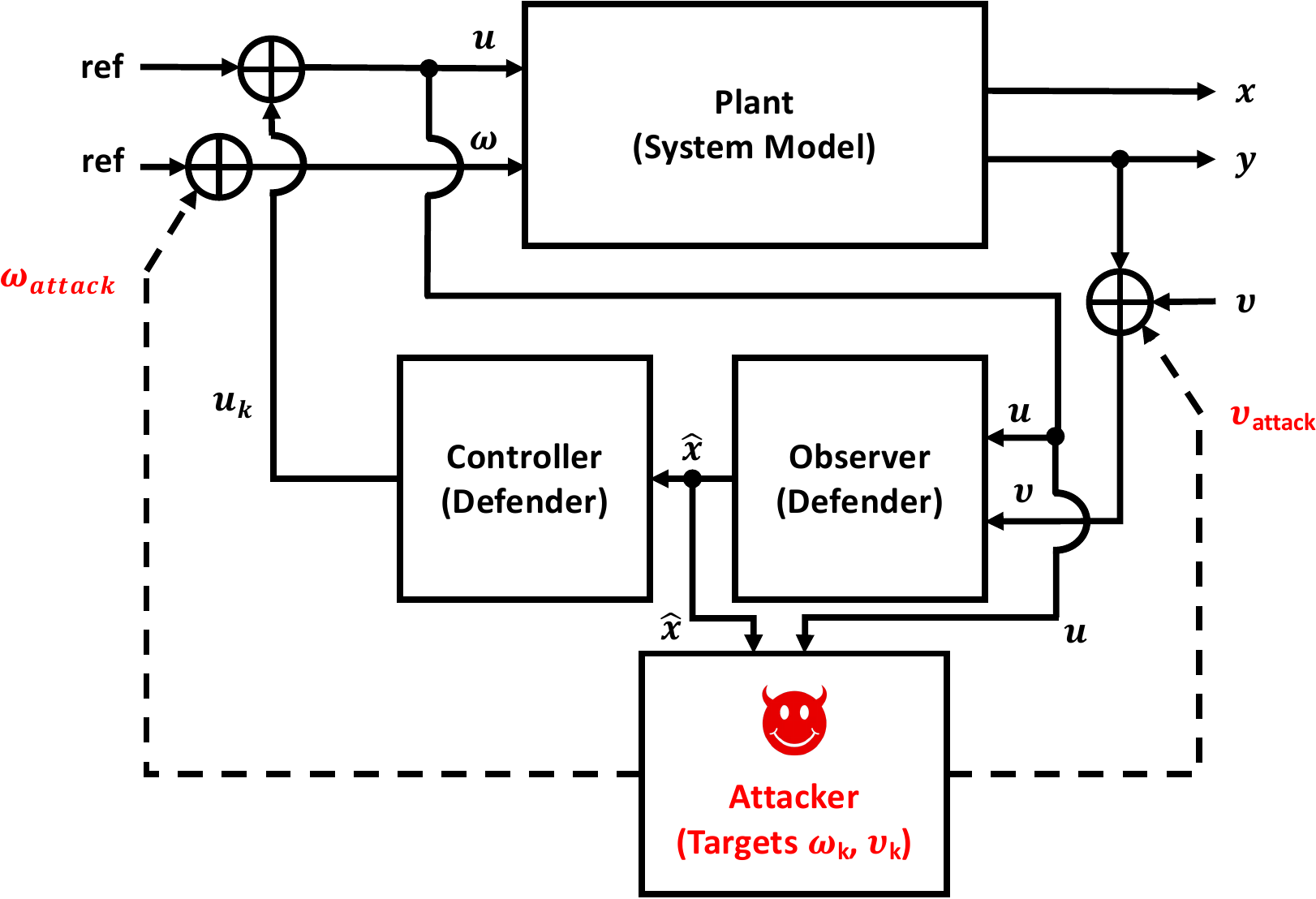}
   \caption{Control system representation of the threat model. Add chi-squared detector to this model} 
  \label{fig:control-sys-model}
  \vspace{-.05in}
\end{figure}

Figure \ref{fig:control-sys-model} depicts a general system structure in which the aforementioned blocks are connected into a system representation.  The goal of an attacker is to encode data into a physical covert channel while maintaining \textit{stealthiness}. To define stealthiness, we first formalize the plant, adversary, and defender models as follows.

\noindent\textbf{Attacker systems model.}  An attacker is formalized to take the plant's estimator output $\hat{x}$ and the controller's output $u_k$ as inputs to its system. In contrast, the defender attempts to monitor (and identify) deviations to the expected control inputs and state. To deviate a system's response, an attacker will add an attack vector to the process noise, $\omega_k$ to the actuators and/or sensor noise, $v_k$ to the measurement, respectively. In doing so, the attacker can break multiple independence assumptions the system state estimator may rely upon for its estimation model. Therefore, the system state, i.e., $x_k$ can now be correlated to the process or measurement noise by the attacker's choosing. The choice and encoding scheme of the attacker will be domain specific and described in the subsequent section. But first, we briefly discuss the defender model in this context.

\noindent\textbf{Defender systems model.} The defender differs from the system's state estimator, in that the defender uses the output of the state estimate and its policy to detect whether a state deviates from its intended path. The goal of the defender will be to distinguish whether a perturbation is due to an attack or merely a random perturbation. This formalization allows us to model the encoding and decoding of data into covert channels, and subject them to systems-theory, when necessary. 
We now use our control-theoretic representation of attacker and defender in the context of covert data exfiltration. We define what an attacker's \textit{stealthiness} and \textit{imperceptibility} is with respect to this systems model.
\begin{definition}[Stealthiness]
We define stealthiness as the attacker's ability to deviate the CPS's state  such that any threshold levels of the system are not crossed as a result of the attack,  the attacked state(s) do not strongly correlate with non-attacked state variables, and the attack is conducted on a state which does not utilize a 'colored-noise' state estimator, i.e., if the measurement noise is correlated, then computing an ensemble average for the auto-correlation of the measurement noise (empirically) would differ from the known correlation signal.

\end{definition}

This basic structure provides an outline for our domain-specific design of both an attack and defense strategy for cyber-physical data exfiltration.

%% file: 4-design.tex
\section{Cyber-physical Data Exfiltration}\label{sec:design}
In this section we formalize the design of a cyber-physical data exfiltration attack over a physical covert channel given the aforementioned system models. The goal of the attacker is to encode data into a physical channel while maintaining stealthiness. The choice of the physical channel and all of the associated parameters will be domain-specific and dependent on the defender model. We will therefore categorize the different attacker-defender scenarios with varying levels of quality for the defender's state estimator.  In all cases, the attacker needs an enhanced sensing modality that can decode the bits at a sufficient granularity. To illustrate each component of the attack design, we present a motivating example of a simplified robotic arm.
\begin{figure*}[htp]
  \centering
  \includegraphics[width=0.9\linewidth]{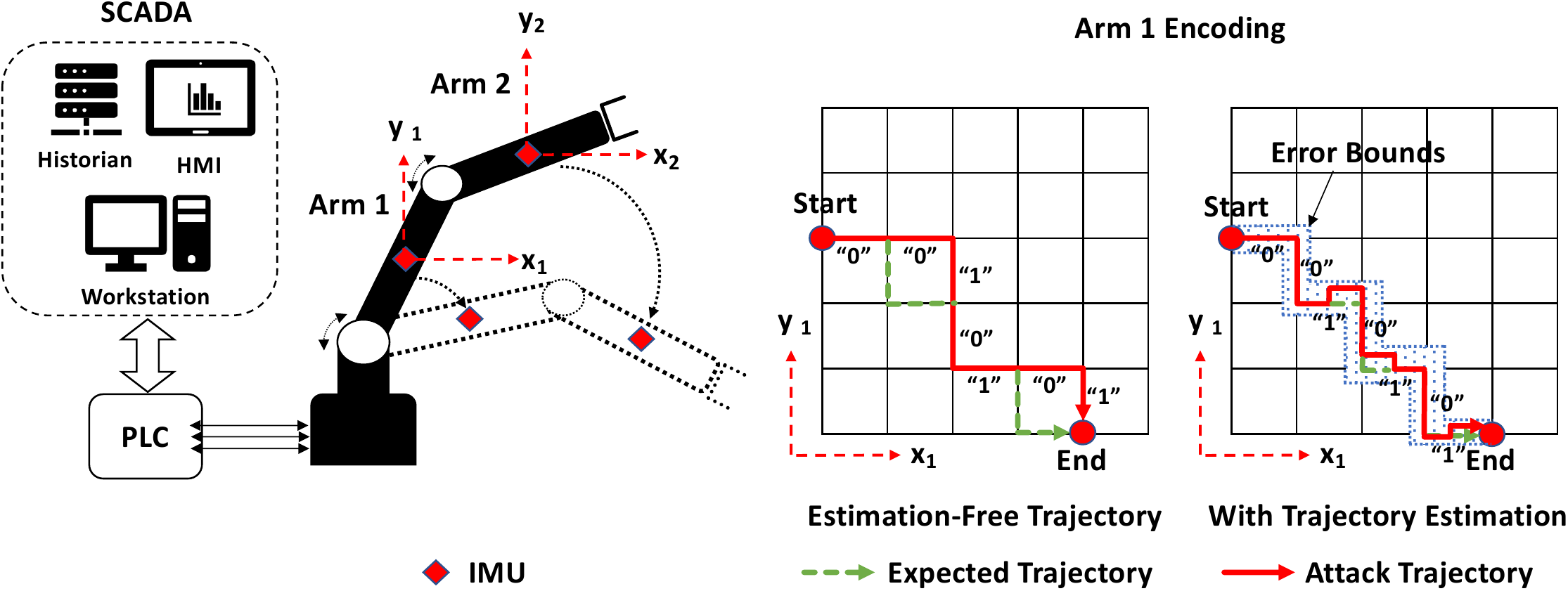}
   \caption{Simplified industrial control system to illustrate cyber-physical data exfiltration. An attacker may encode data into the movement of the robotic arm that is not being estimated or into the noise model associated with movement that \emph{is} being estimated.} 
  \label{fig:simple-robot}

\end{figure*}
\subsection{Motivating Example: Simplified Industrial Control System} 
Figure~\ref{fig:simple-robot} shows an example of a simplified industrial control system (ICS) where a robotic arm is controlled by a programmable logic controller (PLC). The PLC receives higher-level setpoint commands from a supervisory control and data acquisition (SCADA) entity, which may consist of human-machine interfaces, PLC workstations, as well as historians for data logging. The robot arm is composed of two segments that are controlled by stepper motors. In this simplified example, each arm is equipped with an inertial measurement unit (IMU) that is used to close the loop for the arm's controller. For simplicity, we restrict each arm to move along the X-Y plane. In this case, we assume that the PLC takes a an XY-coordinate setpoint from the SCADA entity and its internal control loop calculates the associated actuation commands necessary to arrive at the desired waypoint for both arm segments.  

For this simplified case, we will demonstrate how an attacker may choose a particular attack vector along with the associated parameters. We implemented this industrial control system with a Dobot robotic arm controlled by a Siemens S7 1200 PLC. The robotic arm has a swappable end attachment that can stand in for several example applications, such as 3D printing, laser etching, and a gripper for industrial automation. The PLC controls the arm in a closed feedback loop, reading in sensor data from the arm's two accelerometers and actuating its stepper motors to control the arm's movement. For simplicity, we limit the motion of the arm to a two-dimensional plane. We emulate the SCADA components through an API that sends motion commands to the PLC, which calculates the appropriate actuation commands for the arm's stepper motors.
Given this system, we now describe how a defender would model a state estimator to detect anomalies in the sensor data. 

\noindent\textbf{ICS defender model design.} The typical goal of a defender, e.g., the SCADA entity in this context, is to develop an appropriate state estimator that will detect any anomalies. Theoretically, a perfectly tuned state estimator model for all memory-mapped physical I/O and it's associated physical covert channels of a CPS would detect any cyber-physical data exfiltration attack\cite{bertsekas1995dynamic}. 

In practice, it is difficult to develop a perfectly robust state estimation model for real-world applications as the state dynamics and measurements can be far from ideal. Further, a defender can only develop a state estimation model for the observable set of physical variables--including the physical channels and associated noise models that depend on a particular variable. This means that the state estimation model is heavily dependent on not only the availability and quality of sensors instrumentation, but also the associated level of process noise for the system model. For instance, if the robotic arm is making inferences about it's own pose and reporting just the XY-coordinates of the robots end effector before and after a movement command, then a state estimator is only able to report the posterior $\hat{x}_t$ then prior $\bar{x}_{t+1}$ state estimates and associated covariances of the end effector.
An anomaly could be detected using a distance metric such as a Euclidean distance to see if the current XY-coordinates, $\{(x_1,y_1),(x_2,y_2)\}$ are close enough to the $\hat{x}$ and $\hat{y}$ estimates within a certain error $\epsilon$, i.e.,
\begin{equation}\label{eqn:euclidean}
    \sqrt{(\hat{y_1}-y_1)^2 + (\hat{x_1}-x_1)^2 + (\hat{y_2}-y_2)^2 + (\hat{x_2}-x_2)^2} < \epsilon.
\end{equation}

However, because our system model assumes it is a remote defender, augmenting a defender's state estimator necessitates sending more sensor data over the network, which is contradictory to the edge computation paradigm. In any case, we will detail the design of an attacker's encoding and decoding schemes for varying levels of state estimation.

\subsection{Encoding Data into Physical Channels}
As discussed, an attacker's encoding scheme into a particular physical channel will depend on the quality of the defender's state estimation model--which is assumed to be known by the attacker. As such, we consider three different defender models: (1) an attacker encoding data into a physical channel that is independent from any of the defender's state estimator models, i.e., the control process for that state variable is locally autonomous; (2) an attacker encoding data into the "noise" of a channel that is being directly estimated by the system's state estimator; and (3) the "perfect" defender that is fully aware of an attacker's encoding scheme and is as powerful as the attacker in terms of sensing capabilities. For the latter case, the roles are essentially reversed and the attacker's stealthiness goals will be focused on maintaining confidentiality of the data being encoded. We discuss each case in detail.

\noindent\textbf{Case 1: Local autonomy and estimation.} If a state variable's associated control loop is locally autonomous to the edge device, i.e., there is no feedback control or estimation mechanism from a higher fidelity external entity, then an attacker essentially has little to no inhibitions with respect to stealthiness and can manipulate any aspect of the system as long as the utility of the application is maintained. For instance, in Figure~\ref{fig:simple-robot}, the estimation-free trajectory scenario shows how an attacker may encode data into the path from a starting ("Start") XY-coordinate to an ending ("End") XY-coordinate. Such an attack would need to ensure that the \emph{utility} of the function is maintained, e.g., that the encoding will have a mean noise of zero while ensuring that it reaches a distance within an error bound before the next sample. This also implies that the associated perturbations will not cause any collateral threshold violations for other states being estimated. And although this example shows the path trajectory between two sampled points as the physical channel of choice, any other cyber-physical channel that depends on the associated physical variables can also be utilized by an attacker, e.g., the acoustic noise of the stepper motors during the path trajectory. In any case, the attacker may engineer an encoding scheme that will transmit the data while ensuring the utility function's integrity is maintained. However, encoding data becomes more difficult for subsequent cases where the state variable is being estimated.

\noindent\textbf{Case 2: Remote external feedback control.} If a state variable's associated control loop relies on external state estimation and feedback control, the attacked state variable is being monitored with fine granularity --which complicates the design of the attacker's encoding scheme. However, it is infeasible for a defender to have a perfect state estimator model for real world systems due to environmental and systematic noise. Such an encoding mechanism requires an accurate noise model that is at least as granular as the noise model of the defender. The plot on the right of Figure~\ref{fig:simple-robot} shows an attacker encoding bits into the trajectory of the end effector while staying within an error bounds. In this case, the attacker is much more restricted in terms of how much noise can be introduced in the encoding scheme due to the fact that the bits are being encoded into an estimated variable. However, up until now, we have assumed the attacker is more powerful than the defender. 

\noindent\textbf{Case 3: An omniscient perfect defender.} The final defender model is an ideal "perfect" defender that not only has much more sensing capabilities than the attacker with perfect state estimation, but also knows the attacker model, i.e., the associated encoding and decoding schemes and modalities. In this case, the roles are reversed as an attacker has been exposed and needs to maintain confidentiality of the exfiltrated data. Obviously, a defender could simply take the system offline if the utility of the system is not critical and if it has a means to remotely control the device, i.e., only a subset of the device's remotely controlled variables have been compromised.  But in a honeypot scenario--i.e., where a defender is attempting to discover more information about the attack--the exfiltrated data can reveal the intent of the attacker along with other sensitive semantic information. From a cryptographic perspective, an attacker can leverage two "secrets": (1) standard cryptographic techniques embedded in the encoding software payload and (2) the location and sampling parameters of the decoding sensor.  In this paper, we focus on the latter solution in which the location of a sensor can hide the semantic meaning of data being encoded. Although the design of a cryptographic mechanism within the software payload is outside of the scope of this paper, such an approach has several security and engineering challenges to ensure the semantic information being encoded into the physical actuation is sufficiently secured. If a defender can recover the device, static and dynamic analysis techniques can be used to reverse engineer some of the semantic information, e.g., combining static binary analysis with the dynamic behavioral analysis of the encoder when given certain inputs or environmental conditions~\cite{sun2019mismo}. 

We now present a generalized decoding mechanism for these encoding schemes.

\subsection{Decoding Cyber-physical Encoded Data}
\begin{figure}[htp]
  \centering
  \includegraphics[width=1\linewidth]{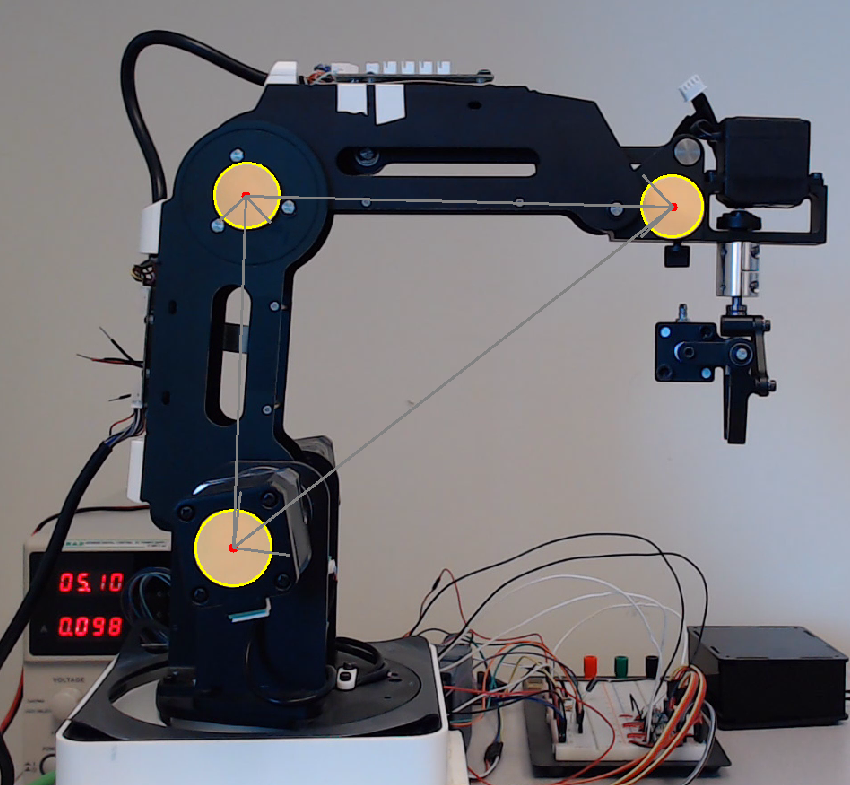}
   \caption{System state tracking using colored markers.} 
  \label{fig:robotarmcv}
  \vspace{-.1in}
\end{figure}

\begin{figure}[htp]
  \centering
  \includegraphics[width=1\linewidth]{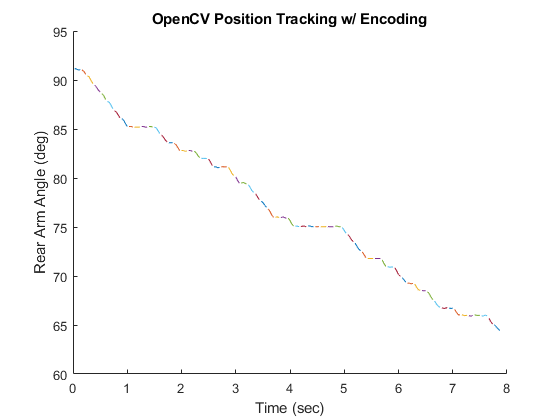}
   \caption{Camera-tracked system state trace. Each segment encodes a single bit based on the change in angle.} 
  \label{fig:robotarm_trace}
  \vspace{-.1in}
\end{figure}

In order to decode data that has been encoded with any of these encoding schemes, the attacker simply needs to mirror the modality and granularity of the encoding scheme. For instance, in the estimation-free trajectory attack of the ICS example, an attacker would need access to the finer-grained path trajectory between movement commands. Having access to either a faster sampling rate or even the IMU data would be ideal, but it is not realistic for a remote attacker--especially if we are assuming the defender does not have access to these results. A more realistic approach is that an attacker may infer the cyber-physical encoding utilizing an air-gapped physical channel such as a microphone monitoring the noise of the device or by visually monitoring the movements of each component from a distance with a camera. For instance, we implemented malicious motion command on the aforementioned PLC that encodes a bit string in the actuation of the arm's motors during a benign motion command. An attacker may focus a camera on the arm to observe specific markers on the arm as shown in Figure \ref{fig:robotarmcv}. We applied color markers to the arm to simplify the tracking algorithm\footnote{A more sophisticated algorithm could perform position tracking without external markers specific to the CPS}. For tracking the markers we utilized OpenCV--an open-source computer vision library.  The resulting output of an encoded movement is shown in Figure \ref{fig:robotarm_trace}. We now discuss the design considerations for a communication protocol given an encoding and decoding scheme.

\subsection{Communication Protocol}
There are several domain-specific design parameters that need to be tuned for particular CPS. Regardless of whether an attacker is using cryptographic mechanisms or not, the goal should be to maximize both the rate of transmission as well as the signal-to-noise ratio (SNR). 

\begin{table}[t]
\centering
\begin{tabular}{|c|c|c|}
\hline
\textbf{Bit Rate} & \textbf{FPS} & \textbf{Bit Error Rate} \\ \hline
5 bit/sec        & 30           & 0\%                  \\ \hline
10 bit/sec        & 30           & 0\%                   \\ \hline
15 bit/sec         & 30           & 15.6\%                     \\ \hline
\end{tabular}
\caption{Bit error rates (BERs) for various encoding rates.}
\label{tab:ber_dobot}
\end{table}
\noindent\textbf{Channel Capacity and Bit Error Rate.} There are several factors that determine the channel capacity of data exfiltration.

\begin{figure*}[!htb]
 \centering
        \includegraphics[width=0.9\linewidth]{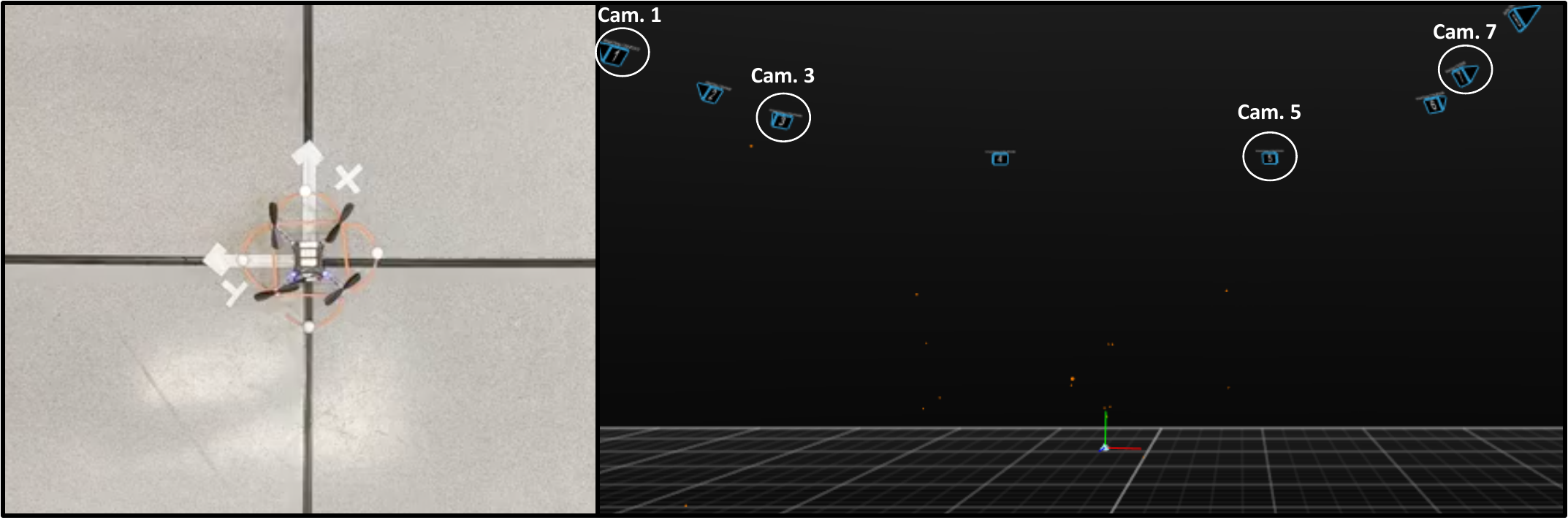} 
        \caption{Drone evaluation setup using the Optitrack motion capture system. The system consists of 12 Optitrack motion capture cameras. The circled cameras were the four different perspectives selected in our evaluation.}
        \label{fig:optitrack-setup}
\end{figure*}

\begin{itemize}
    \item \textbf{Physical system constraints.} The rate of encoding into a physical system is limited by the actuation speed of the system, which is determined by the system's kinematics. Faster encoding speeds require greater forces which the system may not support. Additionally, physical systems may have a minimum precision with which motions can be made consistently (eg. a single motor step is 1.8 for the robotic arm\textdegree).
    
    \item \textbf{The observer's frame rate and resolution.} The channel capacity is also limited by the capabilities of the observer. For a camera, the frame rate is analogous to the sampling rate, and we found that for the robotic arm, at least 3 frames were necessary to identify an encoded motion consistently. Additionally, the resolution of the observer is correlated with the encoding: a higher resolution means that smaller motions can be detected reliably, allowing a greater encoding rate within the constraints of the physical system.
    
    \item \textbf{Maintaining stealth.} In a scenario with a defender performing state estimation on the system, a faster encoding produces more noticeable actuations, increasing the likelihood of revealing the exfiltration process to the defender.
    
\end{itemize}

Table \ref{tab:ber_dobot} shows the bit error rates for decoding in the robotic arm scenario. As we approached the limits of the encoding rate we found that the decoding accuracy decreases significantly due to increased system vibration coupled with fewer frames per encoded bit. We now briefly discuss design considerations for error checking. We now briefly discuss mechanisms that can be utilized to maintain the integrity of the data.

\noindent\textbf{Error checking and redundancy.} Since the transmission channel is a one-way communication link, re-transmission can not be requested in case of a transmission error. Forward error correction such as cyclic-redundancy checks (CRC)~\cite{freivald1999change}, can be used to correct errors at the receiver at the cost of reducing transmission bandwidth for redundancy. Alternatively, if the same variables are being transmitted repeatedly (data values), then the values have a short "lifetime" and we can forgo error correction altogether, filtering out outliers at the receiver. 
The final design piece focuses on an attacker's means of maintaining \textit{imperceptibility} from a defender.


\subsection{Maintaining Imperceptibility} 
The final notion of the aforementioned attacks is maintaining imperceptibility in the face of a ``human observer"--or an observer that may be monitoring the CPS through a particular modality or set of modalities from sensors equivalent to a human's ``sensors". This problem is analogous to the problem of adversarial machine learning where an attacker is introducing perturbations to a model's input data while minimizing some loss function such that the system will misclassify the data sample while maintaining imperceptibility of the perturbations~\cite{papernot2016limitations}. In this context, the perfect model of perceptibly is the associated decoding mechanism itself. In addition to maintaining stealthiness with respect to the state estimation model, an attacker will also minimize the encoded movements such that the decoding function will only work with a decoder that has a sufficient sensing granularity, e.g., a camera equipped with an appropriate focal length to pickup tiny movements of the robotic arm.

We propose the following simple scheme for maintaining imperceptibility. For a given attacker strength, it is desirable to encode information at the lowest SNR that the attacker can still decode reliably (eg. with an acceptable bit error rate). By definition, this minimizes the differentiation between signal and noise for any observer and results in the least conspicuous encoding. Additionally, the frequency and choice of encoding should be chosen carefully to closely mirror normal operating characteristics. That being said, determining these parameters may be impractical in certain situations.
We now evaluate each of the aforementioned attacker-defender scenarios on a much more complex autonomous edge device.

%% file: 5-evaluation.tex
\section{Evaluation on an Autonomous Edge Device}\label{sec:eval}

We now evaluate both the attacker and defender models presented from the previous section in the context of a more complex edge computation scenario: a surveillance drone. We are emulating the aforementioned scenario depicted in Figure~\ref{fig:system-model} where a drone is part of an IoT coalition supporting a group of first responders or soldiers. 
In particular, the drone is tasked to surveil an area, e.g., to search for particular objects of interest. Any inferences made by the drone will be reported by back to a supervisory entity that is interacting with the drone. We describe our experimental setup in detail. 

\begin{figure}[htp]
  \centering
  \includegraphics[width=1\linewidth]{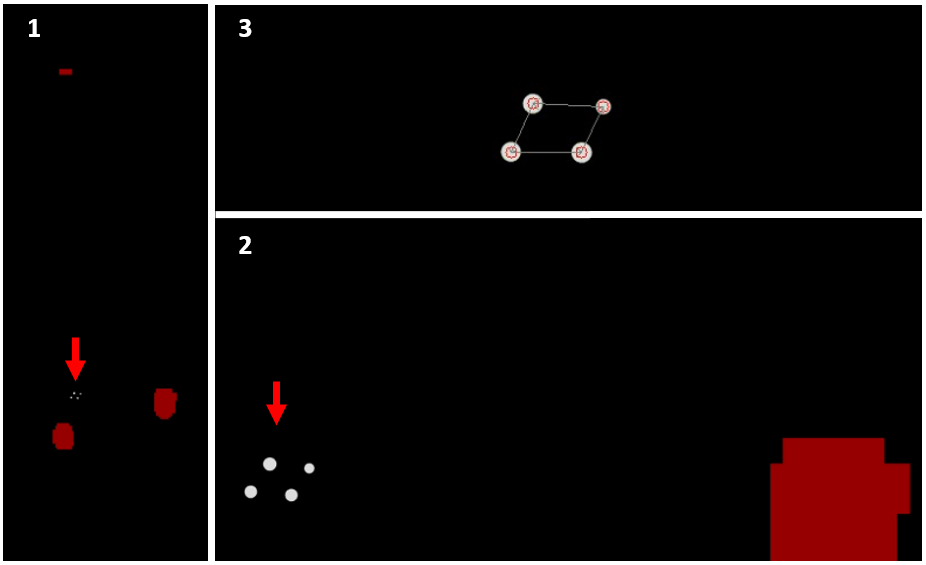}
   \caption{Optitrack replays used to simulate an enhanced observer: 1. Normal, 2. Enhanced, 3. Tracking example.}
  \label{fig:drone_zoom}
  \vspace{-.1in}
\end{figure}

\subsection{Experimental Setup}

To evaluate the drone surveillance application, we formalize our defense models on both position and orientation state estimates of a Crazyflie quadcopter~\cite{giernacki2017crazyflie}. For our scenario, we abstract the specific task and focus on the trajectory of the drone since the Crazyflie cannot support such a large computation load\footnote{Although the Crazyflie cannot support such large computation, neural accelerators have already shown such inferencing can be run on the edge. It is a safe assumption that this computation is already or will be soon enabled on larger outdoor drones.}. Throughout the trajectory, the ideal attacker can choose to deviate any physical degree of freedom of the system such as position, speed, or orientation. For simplicity, we show the effect a defender can have when an attacker is physically exfiltrating data through the drone's yaw variable. We use an Optitrack motion capture system to provide the drone with external location estimates of its 3D position and orientation in space.  The Crazyflie was fitted with four Optitrack markers  to precisely localize the drone during its flight. Our Optitrack setup utilizes 12 cameras to obtain sub-millimeter positioning accuracy. This state of the art level of accuracy allows us to provide the defender with a very precise and accurate state estimator for the drone--more precise than outdoor localization schemes\footnote{The Crazyflie quadcopter was chosen because it can be flown indoors and the motion capture system needs to be calibrated in a static indoor environment.}. The drone with the markers as well as its representation in the Optitrack software can be seen in Figure~\ref{fig:optitrack-setup}. We use the Robotic Operating System (ROS) as the software package to communicate between the drone, motion capture system, and host computer in real time. 

We evaluate a defender model under the three aforementioned attack cases for our experimental setup. For the first case, the drone is tasked to execute a constant hover at 0.5 meters. The attacker encodes the data into the yaw variable about a fixed position. For the second case, we allow the defender to monitor both the position and yaw variable of the drone. Because the defender would be able to easily detect a change in yaw for a stationary hovering case, the drone is tasked to fly in a circle approximately 1 meter radius at 0.5 meters from ground level. In the third case, the drone is tasked to hover again, but now the attacker exfiltrates data with an asynchronous and more challenging encoding scheme to highlight alternative means of stealthiness in the face of a perfect defender.

The attacker generally uses two encoding schemes: The first scheme is used in scenarios one and two.
\begin{itemize}
    \item \textbf{Encoded Bit 1} Attacker yaws drone approximately 5 degrees counter-clockwise from start. Attacker yaws back to reference to complete the transmission.

    \item \textbf{Encoded Bit 0} Attacker yaws drone approximately 5 degrees clockwise. Attacker yaws back to reference to complete the transmission.
\end{itemize}

To evaluate the feasibility of decoding the cyber-physical encoded data, we utilized video recordings of replays from the Optitrack system. The recordings were taken at 30 frames per second, and zoomed-in simulates an enhanced observer as shown in Figure \ref{fig:drone_zoom}. Similar to the ICS scenario, relative angles were established between the markers to track the system state for data exfiltration.

\begin{figure}[htp]
  \centering
  \includegraphics[width=0.99\linewidth]{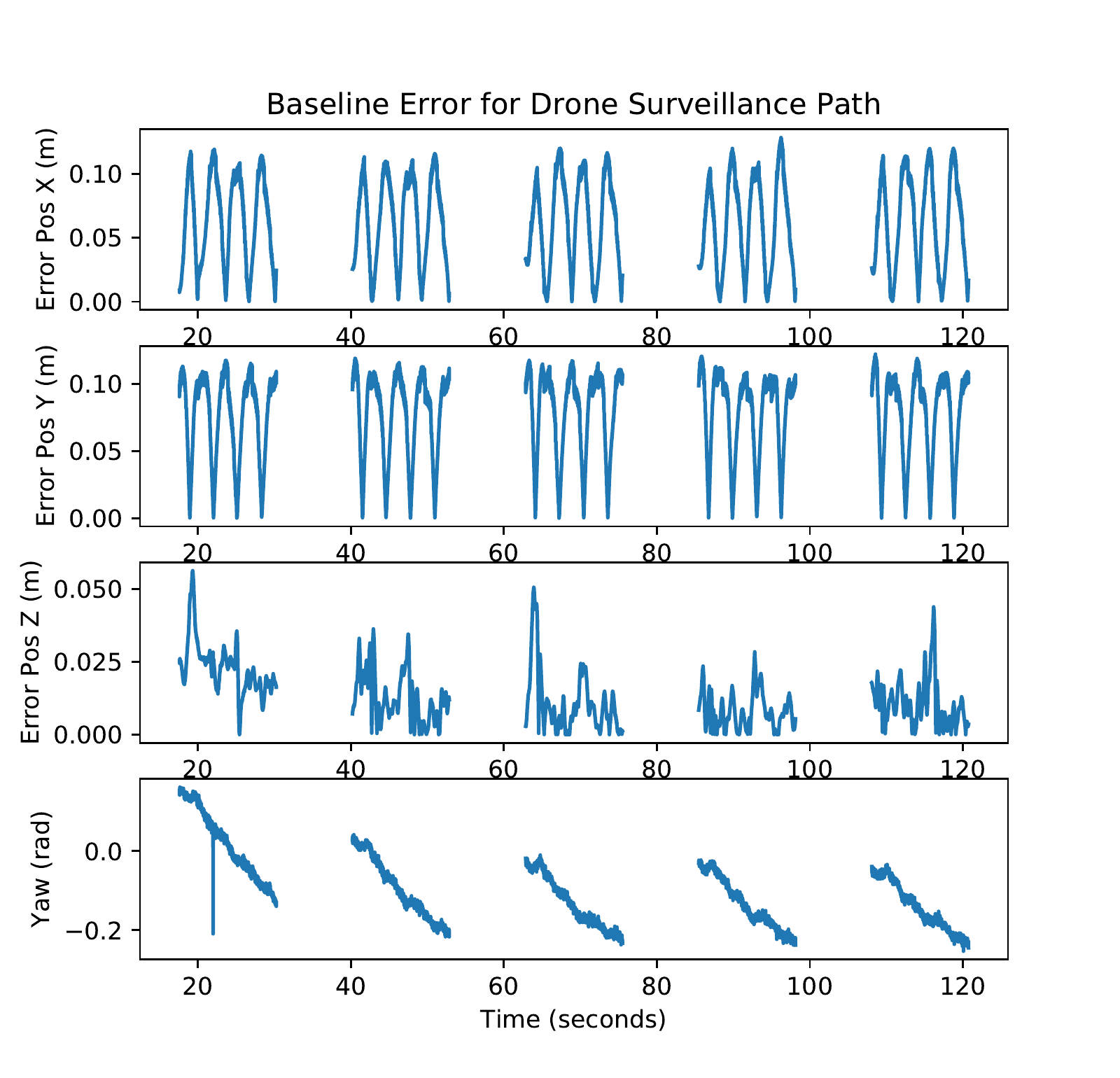}
   \caption{Baseline errors for drone surveillance path.}
  \label{fig:drone_yaw_pos_ref}
  \vspace{-.1in}
\end{figure}
\begin{figure}[htp]
  \centering
  \includegraphics[width=0.99\linewidth]{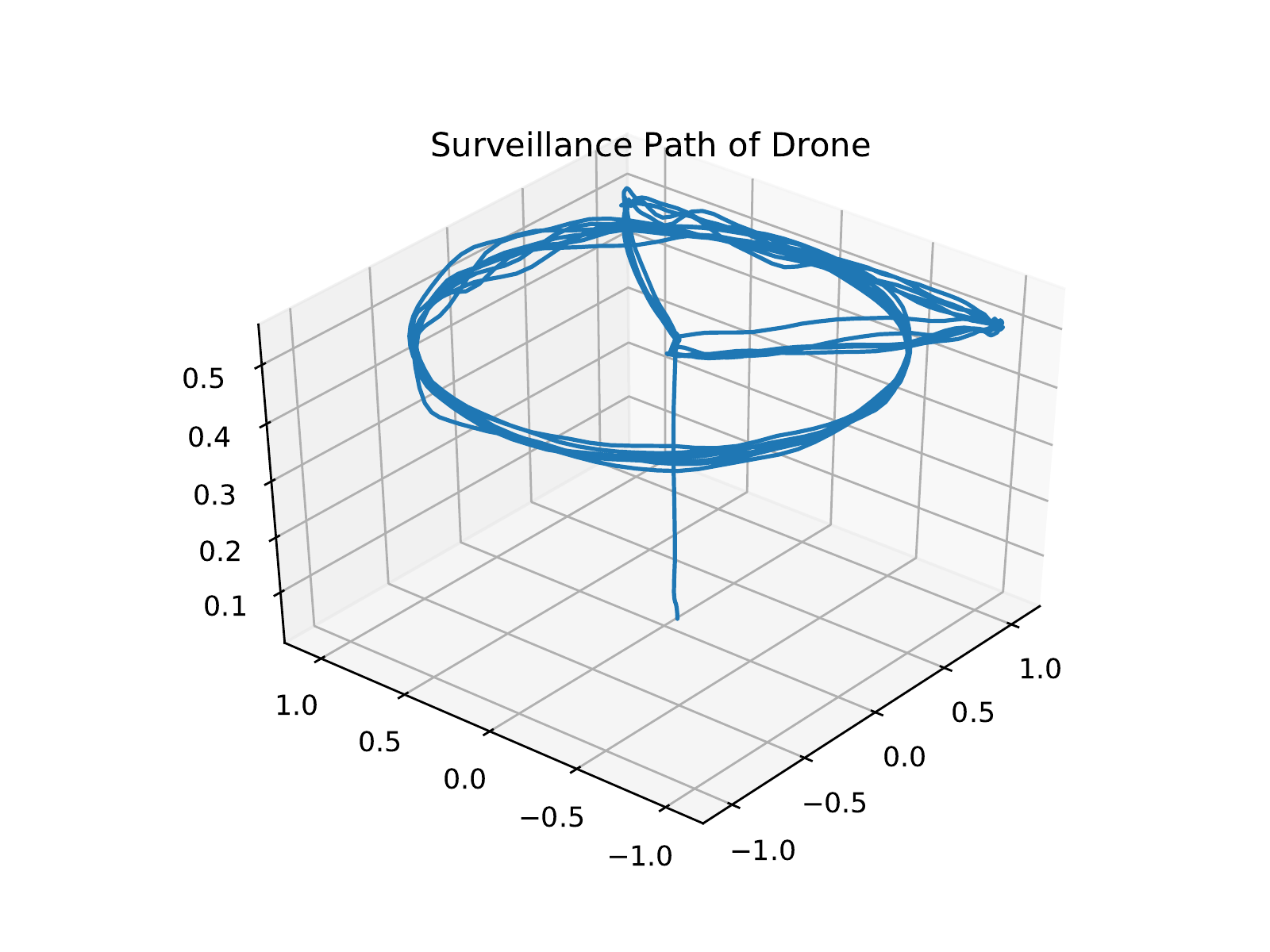}
   \caption{Baseline surveillance loop of drone.\luis{We need to include a chi-squared detector for the anonamly detector}}
  \label{fig:baseline-loop-3D}
  \vspace{-.1in}
\end{figure}

Baselines with no attacker perturbation were first found for both the hover and circle scenarios. For the hover scenario, ambient noise levels were seen to be small. For the circle (surveillance) scenario, the drone conducted ten circles with no attacker perturbation for ground truth. Figure~\ref{fig:baseline-loop-3D} depicts the surveillance loop of the drone, and Figure \ref{fig:drone_yaw_pos_ref} shows the baseline position error and yaw of the drone as it completes 10 circles with no attacker perturbation. Furthermore, Figure \ref{fig:drone_yaw_pos_ref} details five segments per each subplot, depicting the respective amount of position error or yaw as the drone navigates two full circle paths, resets and starts again. We observe the XYZ position error variances are .00123, .00101, and .0001 $m^2$, for the case of no attacker, respectively. 


\subsection{Evaluating Across Different Defense Schemes}
We now provide an evaluation of different attacks across the aforementioned defender models.
 
\begin{figure}[tp]
  \centering
  \includegraphics[width=0.99\linewidth]{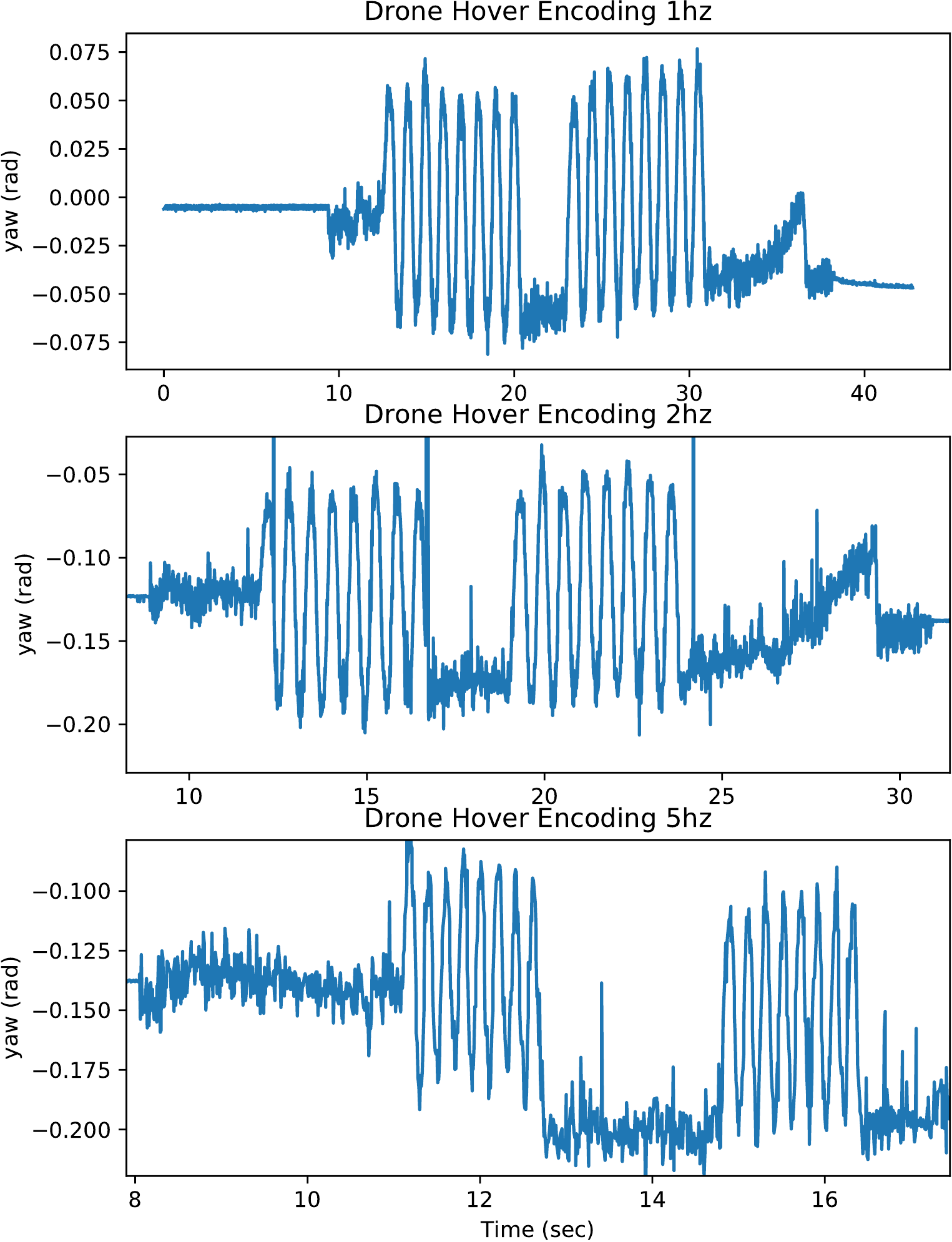}
   \caption{From Top: Drone yaw (radians) vs flight time (seconds) for exfiltrating a 1 and 0 at about 1hz, 2hz and 5hz, respectively} 
  \label{fig:drone_hover_freqs}
  \vspace{-.1in}
\end{figure}

\noindent\textbf{Case 1: Local autonomy and estimation.}
For case 1, the attacker is encoding into the yaw variable that is not being estimated by the defender. The attacker exfiltrates using the first bit encoding scheme. We repeat this process for encoding speeds of 1 bit/s, 2 bit/s and 5 bit/s. Figure \ref{fig:drone_hover_freqs} illustrates the hovering sequences of the drone. The data presented in Figure \ref{fig:drone_hover_freqs} furthermore shows the yaw of the drone does not undergo significant drift as the attacker perturbs the system. Simple threshold values are sufficient for the defender to detect an attacker in this experimental setup. Table \ref{tab:hover_encode_cont} summarizes these results.
 
\begin{table}[bp]
\centering
\begin{tabular}{|c|c|c|c|c|}
\hline
\textbf{Freq} & \textbf{Mean} & \textbf{Thresh Low} & \textbf{Thresh High} & \textbf{Accuracy}\\ \hline
1Hz  & 0 & -.025  & .025  & 100\%                  \\ \hline
2Hz & -.125  & -.035 & .035 & 93.75\%                \\ \hline
5Hz   & -0.14   & -.030  & 0.30 & 93.75 \% \\ \hline
\end{tabular}
\caption{Defender Results for Case 1. Defender has knowledge of attacker encoding strategy. Defender uses threshold detection levels to detect attack\matt{visualize and contextualize ($\chi^2$) better}}
\label{tab:hover_encode_cont}
\end{table}

\begin{figure}[tp]
  \centering
  \includegraphics[width=1\linewidth]{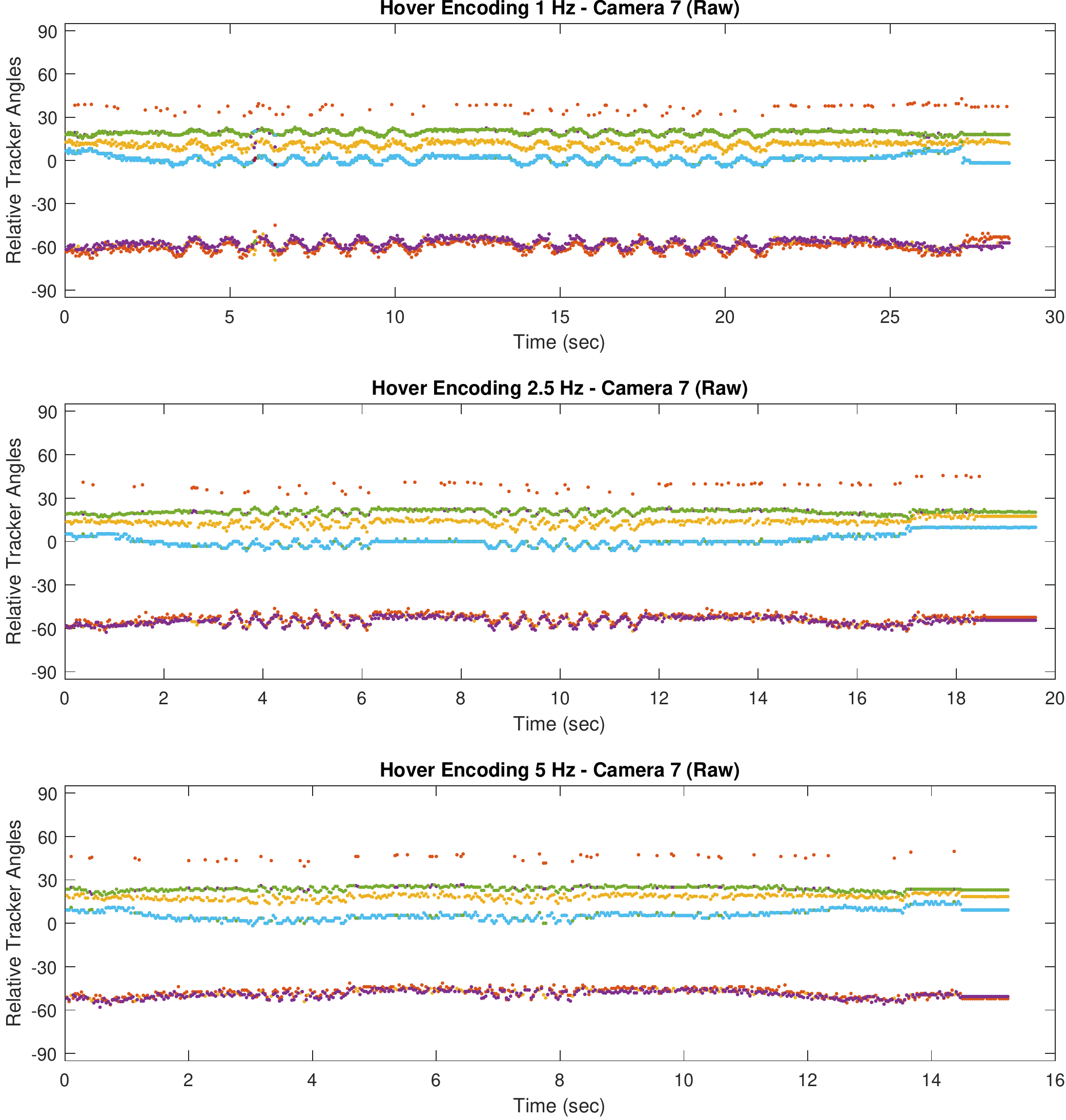}
   \caption{Visually reconstructed drone encoding trace.}
  \label{fig:drone_cv_trace}
  \vspace{-.1in}
\end{figure}

\begin{figure}[tp]
  \centering
  \includegraphics[width=1\linewidth]{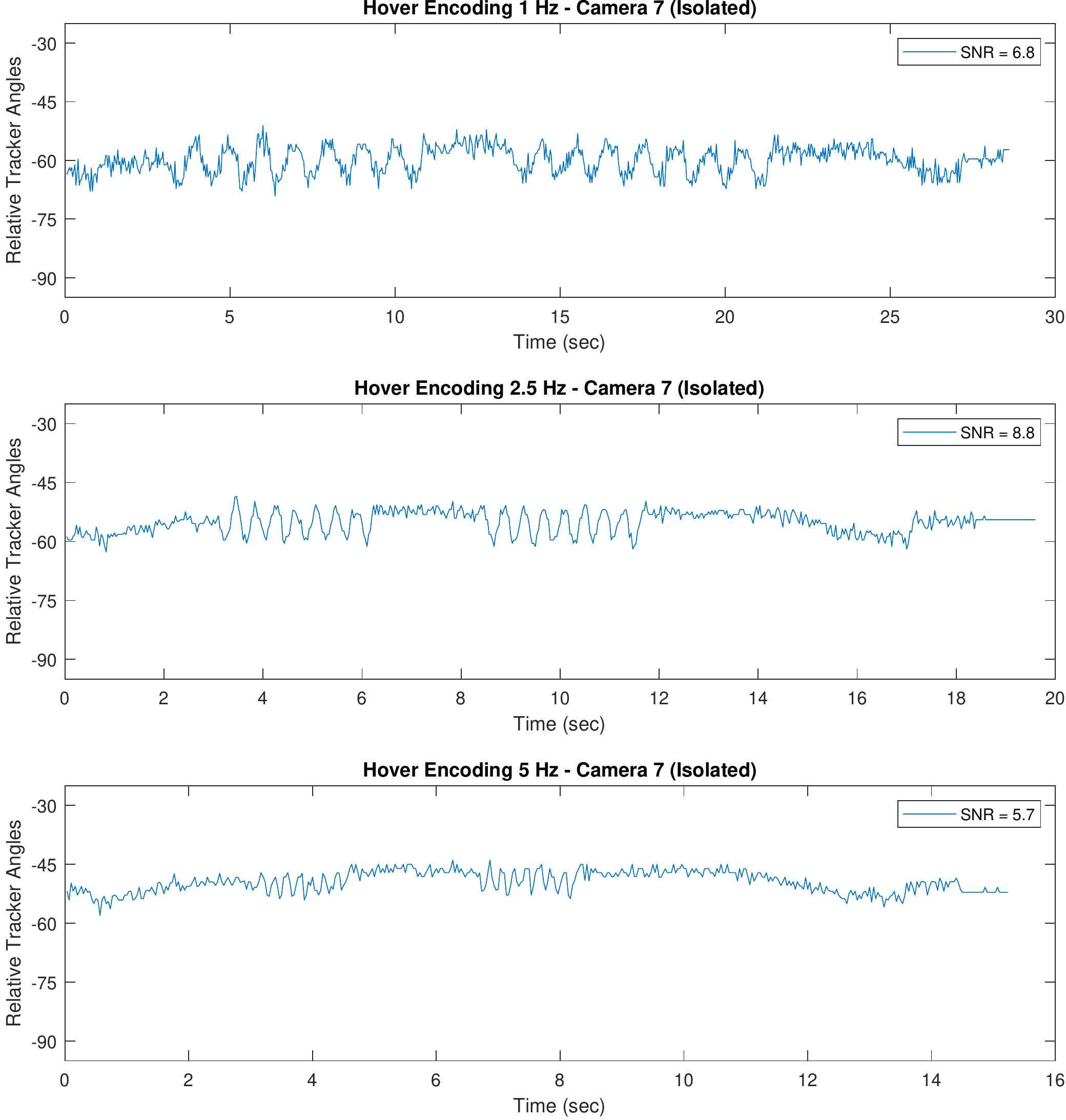}
   \caption{Isolated encoding trace for decoding.}
  \label{fig:drone_cv_trace_single}
  \vspace{-.1in}
\end{figure}

From the attacker's perspective, observing the drone's motion from afar provides a reconstruction shown in Figure \ref{fig:drone_cv_trace}. Noisy artifacts are present due to flickering markers and temporary occlusions. Isolating a single channel in Figure \ref{fig:drone_cv_trace_single} reveals a signal with acceptable signal-to-noise ratios (dB) for decoding (with sufficient signal processing). Of note in Figure \ref{fig:drone_cv_trace_single}, as the encoding frequency approaches the channel capacity, physical system constraints become apparent, as the drone must either endure greater accelerations or make smaller rotations (observed) to maintain the 5 Hz bitrate.

\begin{figure}[htp]
  \centering
  \includegraphics[width=0.99\linewidth]{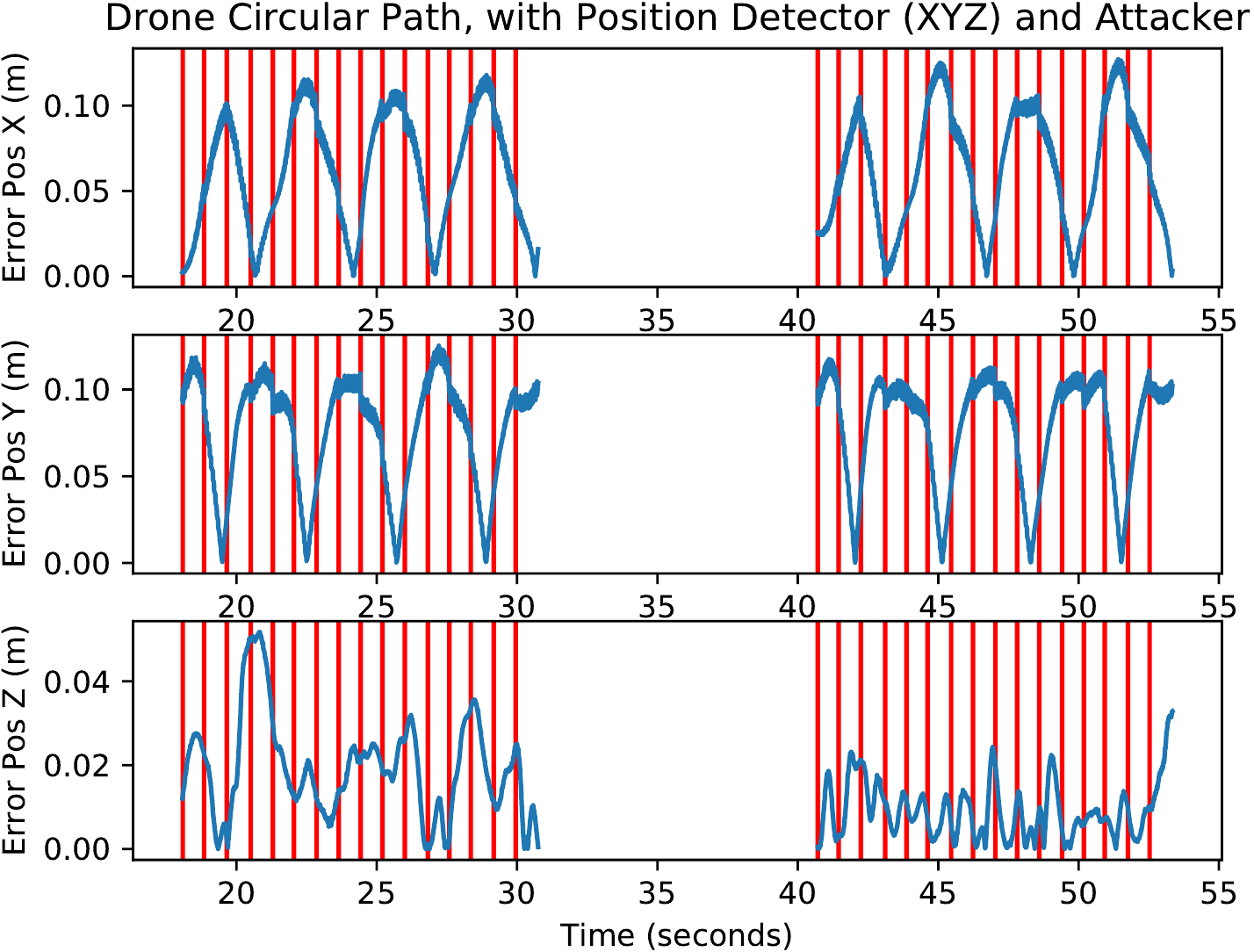}
   \caption{Error in XYZ Position as attacker encodes bit sequence through yaw. Encoding is unobservable when monitoring just position. Start of attacker's bit encoding represented in red. \matt{augment with error thresholds (not violated)}}
  \label{fig:drone_pos_att}
  \vspace{-.1in}
\end{figure}

\noindent\textbf{Case 2: Remote external feedback control.}
For case two, the attacker is encoding into the yaw variable as the drone flies along its circular surveillance path. Figure \ref{fig:drone_pos_att} shows the 3D position error of the CPS under encoding scheme 1. The attacker's encoding begins at each vertical red line, respectively. In this example, we first give the defender the sub-task of monitoring the position of the drone. Figure \ref{fig:drone_pos_att} shows the position error traces as the drone follows the circular reference path. One key importance of Figure \ref{fig:drone_pos_att} is the fact that the start of the attacker's encoding signal does not influence error in 3D position. This is further evident when comparing the associated xyz position error variance levels from the attacker case (.00116, .00098, .000118) $m^2$, respectively. Thus, we observe the yaw-attack has no effect on the position error variance levels and the attack is unobservable. 

\begin{figure}[htp]
  \centering
  \includegraphics[width=0.99\linewidth]{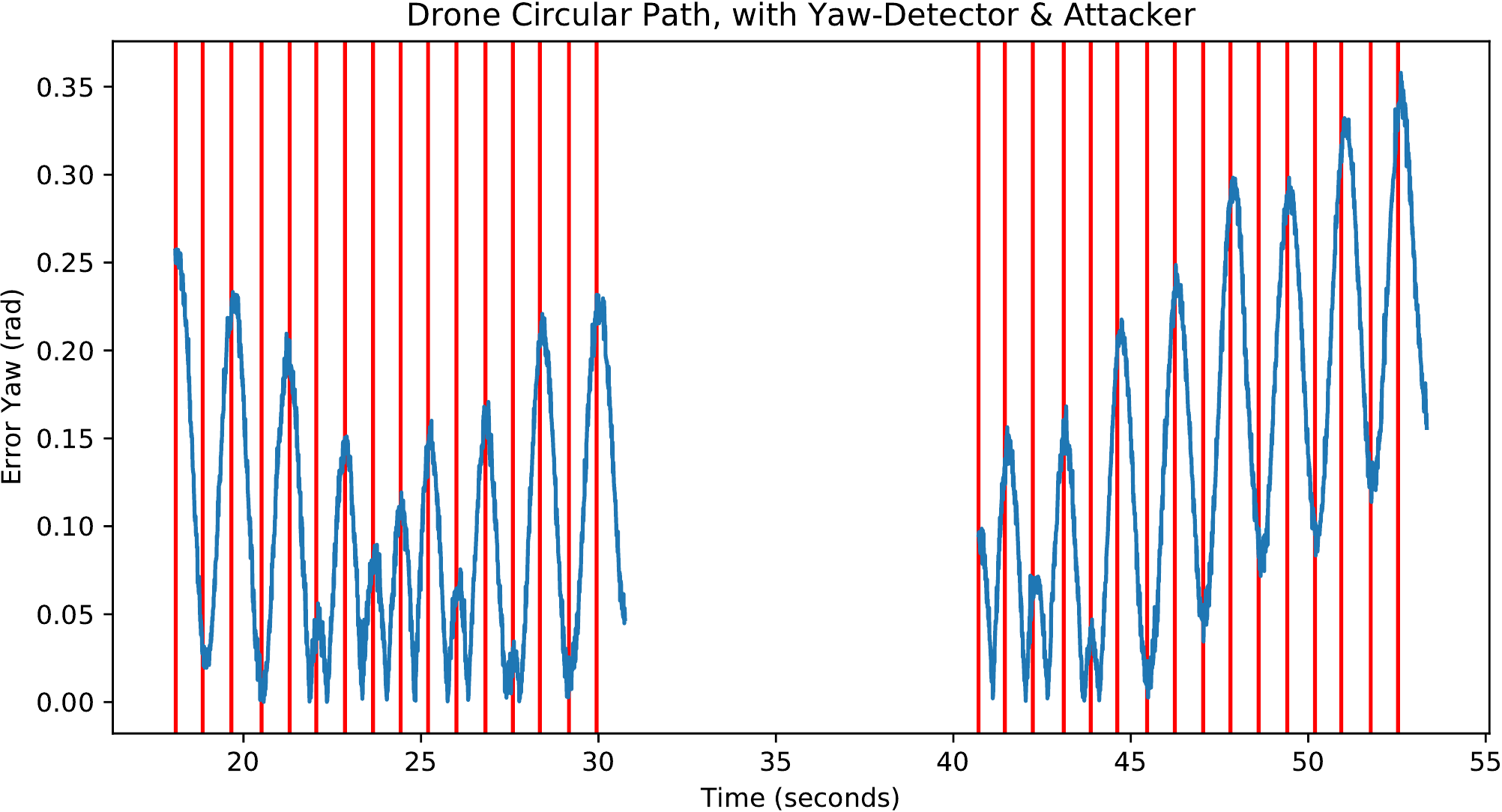}
   \caption{Error in yaw (radians) vs flight time (seconds). Error in yaw is in sync with the start of each attacker perturbation command (red lines)} 
  \label{fig:drone_yaw_att}
  \vspace{-.1in}
\end{figure}

Figure \ref{fig:drone_yaw_att} depicts the yaw error about the reference path of the drone as the attacker performs encoding scheme one again. In this case, the defender solely monitors the yaw variable, while the attacker perturbs this channel. The start of each attacker encoding is represented by a vertical red line. We see that maximums and minimums of yaw directly align with the attacker's encoding frequency over this channel. 

As seen from Figure \ref{fig:drone_yaw_att}, the drone's yaw slightly drifts as the attacker perturbs the drone's heading during its circular flight. We see implementing a simple thresholding technique here will not be as robust when compared to the hover case, since the flight data is clearly non-stationary in this example. Given the defender is monitoring the yaw state variable, and knows the attacker's bit encoding scheme, a simple local min/max extrema search would detect the attacker's presence and encoded bit sequence. Table \ref{tab:circle_encode} displays the defender's accuracy in correctly detecting the exfiltrated data through thresholding and local extrema finding.

\begin{table}[bp]
\centering
\begin{tabular}{|c|c|}
\hline
\textbf{Technique} & \textbf{Accuracy}
\\ \hline
Local Extrema  & 90.6\%                  
\\ \hline
Thresholding &  53.1\%                
\\ \hline
\end{tabular}
\caption{Defender Results for Case 2: Defender has knowledge of attacker encoding strategy. Comparison of threshold detection and local extrema accuracy.}
\label{tab:circle_encode}
\end{table}

\begin{figure}[htp]
  \centering
  \includegraphics[width=1\linewidth]{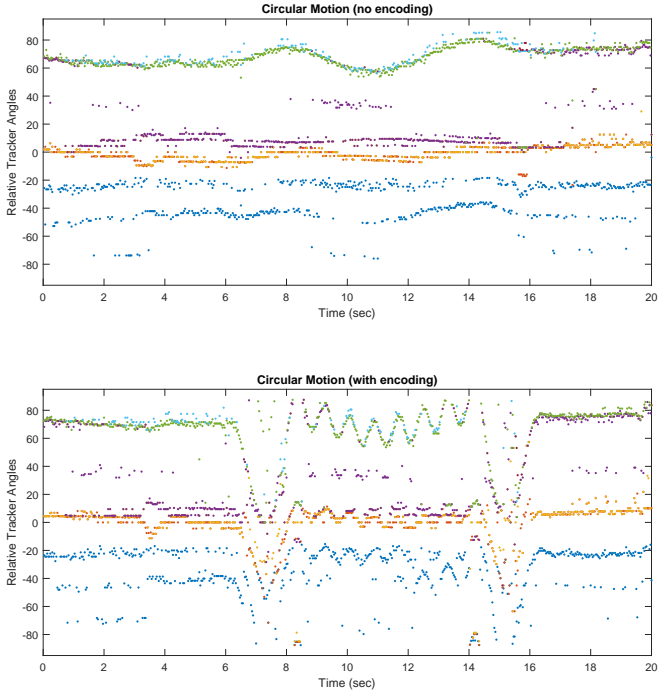}
   \caption{Visually reconstructed circular motion trace.}
  \label{fig:drone_cv_trace_circle}
  \vspace{-.1in}
\end{figure}

For the circular flight scenario, it is again possible for an attacker to reconstruct and decode encoded data as the yaw is independent of the XYZ position. In this situation the attacker faces challenges similar to the defender when decoding (such as non-stationary data), as both parties are estimating the yaw. However, an unencoded flight can help establish a baseline for decoding. Figure \ref{fig:drone_cv_trace_circle} shows the adversarial observer's reconstruction of the circular motion. Although the defender in this case would also be able to read and decode encoded data, we discuss in following sections ways to protect an attacker and their communication.

\begin{figure}[htp]
  \centering
  \includegraphics[width=1\linewidth]{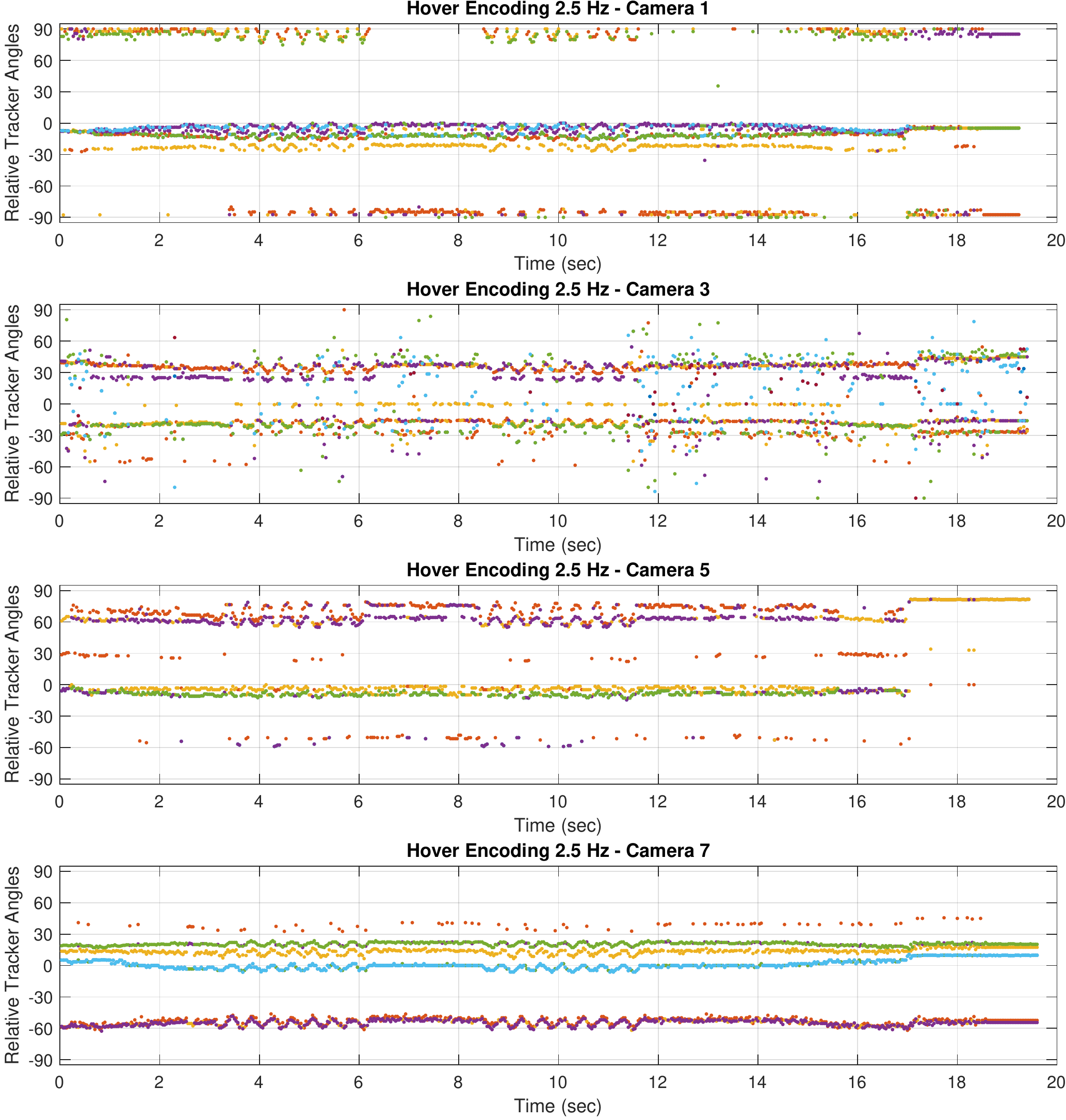}
   \caption{Decoding as seen from various angles (see Figure \ref{fig:optitrack-setup} for camera locations).}
  \label{fig:drone_cv_angles}
  \vspace{-.1in}
\end{figure}

\noindent\textbf{Case 3: An omniscient perfect defender.}
In the case of a perfect defender, who knows both the attacker's covert physical channel and encoding scheme, there are two main secrets for the attacker to protect:
\begin{itemize}
	\item \textbf{The communication contents.} In the case of an ideal defender, the roles effectively swap -- the defender wants to find out what information is being exfiltrated, whilst the attacker must defend this secret. To prevent a defender with knowledge of the system from figuring out what data the attacker is exfiltrating, encryption (based on some shared secret between the compromised device and the attacker) can be used, which is outside of the scope of the paper.
	\item \textbf{The location of the decoding sensor.} Depending on the encoding channel and scheme, it is possible that data exfiltration is only feasible from certain perspectives (eg. the 2D robotic arm scenario). When the defender has this information, it may compromise the stealth of the attacker by giving away their location.
	Given the rotational symmetry of the yaw in the drone case, we hypothesize that this encoding channel gives little information on where the attacker is observing the system from. Figure \ref{fig:drone_cv_angles} shows that this is indeed true; the encoding is visible from multiple camera angles. It is important to note though that certain angles are still "better" than others, whether due to occlusion or a observation angle depth leading to a larger SNR. However, given a capable attacker all of these angles are sufficient for data exfiltration without giving information to a knowledgeable defender.
\end{itemize}

\noindent\textbf{Obfuscated Encodings:}
We now consider the case when the attacker exfiltrates a meaningful byte of data from the drone. 

A second, more challenging encoding scheme is now used:
\begin{itemize}
    \item \textbf{Encoded Bit 1} Attacker yaws drone approximately 5 degrees counter-clockwise. Attacker holds this orientation for 0.75 seconds to transmit another encoded bit 1.

    \item \textbf{Encoded Bit 0} Attacker yaws drone approximately 5 degrees clockwise. Attacker holds this orientation for 0.75 seconds to transmit another encoded bit 0.
\end{itemize}

\begin{figure}[htp]
  \centering
  \includegraphics[width=0.99\linewidth]{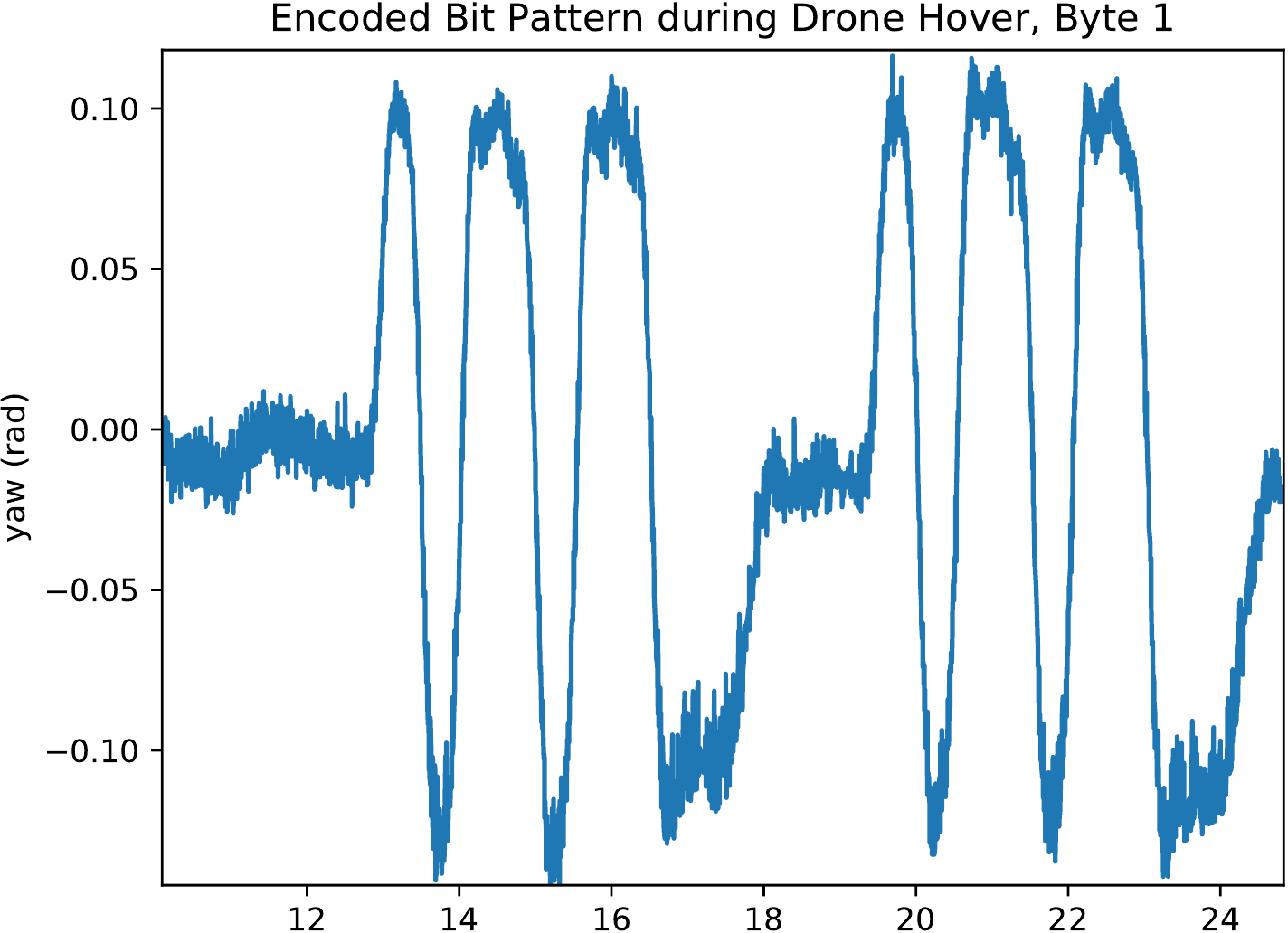}
   \caption{Attacker exfiltrates a byte of data by yawing the drone with precise timing to exfiltrate bits covertly. Byte 1 is: 10110110} 
  \label{fig:drone_byte_encode}
  \vspace{-.1in}
\end{figure}

Figure \ref{fig:drone_byte_encode} depicts a byte of data being physically exfiltrated through the drone's yaw using the encoding scheme above. The byte is transmitted twice, to show reproducibility of the transmission. In this case, the attacker's exfiltration strategy is dependent on both direction of physical perturbation and the duration of the perturbation. If the defender chooses to implement either of the two formerly proposed techniques (thresholding, local-extrema search), the attacker will have successfully fooled the defender in recovering the wrong bit-string. For example, the byte exfiltrated from figure \ref{fig:drone_byte_encode} is 10110110, however, the defender would guess 101010 if they only had knowledge of the attacker's perturbation rule and not the time delay rule. This demonstrates the attacker can fool the defender by encoding bits over both the system's degrees of freedom and time. This situation is further exacerbated by the fact the attacker may be using cryptographic mechanisms to communicate the exfiltrated bit string to a third party observer. We hypothesize combining secure state estimation models to estimate the current attack vector and possibly time-series machine learning models to infer the exfiltrated bits from the attack sequence could provide a solution for the asynchronous exfiltration scenario described here, and will be part of future work.

%% file: 6-related-work.tex
\section{Related Work}\label{sec:related}
In this section we will discuss some of the related work on data exfiltration via covert channels as well as the formalization of side channels.

\noindent\textbf{Air-gapped covert data exfiltration.} There is a large body of research on the topic of physical covert channel data exfiltration across air-gapped systems. Multiple works have shown that electromagnetic signals emitted from devices, e.g., signals from video displays~\cite{guri2014airhopper}, GSM frequencies emitted from workstations~\cite{guri2015gsmem}, or USB generated electromagnetic emissions~\cite{guri2016usbee}, can be picked up by mobile phones to establish a physical covert channel. It was shown that even if these channels were physically insulated to conceal any emissions via a Faraday cage, magnetic fields emitted from a CPU can act as a transmitter of data to mobile phones~\cite{guri2018odini,guri2018magneto}. Power consumption has also been utilized as a transmitter of data by modulating the CPU utilization~\cite{guri2018powerhammer}. Similarly (but at a larger scale), it was shown that two PLCs in the context of industrial power grid can communicate covertly with each other by modulating their associated actuators in a stealthy manner~\cite{garcia2014covert}. However, these systems are not necessarily air-gapped as they have direct access to the cyber-physical sensors.  Thermal emissions between two PCs have also been utilized to establish bi-directional communication~\cite{guri2015bitwhisper}. Acoustic covert channels have been utilized to exfiltrate data from physical hard drive noises~\cite{guri2017acoustic}, commodity desktop speakers~\cite{guri2018mosquito}, or desktop fans~\cite{guri2016fansmitter}. It has even been shown propietary information of 3D printed models can be divulged from the noise of the motors~\cite{faruque2016acoustic}. There has also been several works that have shown a similar approach to optically encode and decode information by utilizing LEDs~\cite{guri2017led,gur2018xled,guri2019air}. In all of these cases, these attacks were presented informally and, to the best of our knowledge, our work is the first to formalize a control-theoretic model of covert physical channel exfiltration while maintaining \textit{stealthiness} as well as the utility of the respective cyber-physical application. Amost all of these related works actually do not implement these attacks in the context of cyber-physical applications and the associated proposed countermeasures discuss physical isolation or procedural security that are not applicable to cyber-physical edge devices in the wild.  It is important to note that there have been attempts to formalize the notion of \textit{side-channels}.

    
\noindent\textbf{Formalizing cyber-physical side channel attacks.} The notion of an information-theoretic model for side-channels has been discussed to describe what an attacker may derive from other types of side channels. In these attacks, an attacker can query a system to observe its characteristics and infer characteristics about a secret key given a limited number of queries~\cite{kopf2007informationf}. Note that the analysis of side-channels are subsumed by our physical covert channel analysis as side-channels are cyber-physical dependencies stemming from the memory-mapped I/O of the system. An attacker or a defender can utilize our approach to analyze the cyber-physical dependencies of memory-mapped I/O and uncover possible side-channels that may leak information. The major difference is that our model assumes an attacker can compromise the CPS binary to encode information and instrument side-channels as covert channels.

%% file: 7-discussion.tex
\section{Discussion}
We briefly discuss the practicality of such attacks as well as the practical design of defensive countermeasures.

\noindent\textbf{Practicality and efficacy of attacks.} The attacks presented in this paper are significantly more complicated than the previous air-gapped attacks, e.g., encoding data into LEDs is much easier than encoding into the movement of a drone while maintaining the utility of the application. However, such attacks are also easier to mitigate as one can simply physically disable such unnecessary actuators to harden the systems, e.g., by removing any LEDs from the system. It is much less feasible to constrain particular movements of a CPS. Further, the attacks presented in this paper were very simple movements. For more complex systems with more physical degrees of freedom, e.g., a swarm of drones or a factory automation floor with several robotic arms, more sophisticated encoding and decoding mechanisms can be instrumented to both increase the rate of data transmission as well as to further obfuscate the encoding scheme.

\noindent\textbf{Practical defensive measures.} The state estimators utilized by the defenders in this paper were idealistic as we used the Optitrack motion capture system that has sub-millimeter accuracy and typically requires significant calibration for a small and limited space. In reality, localization and state estimation of drones in the wild is much more noisier and less predictable. In such cases, state estimation may not be reliable enough to detect an attacker and a defender may need to rely on a means of attesting the software that is running on the CPS. Recent works have made strides towards attesting the behavioral integrity ~\cite{adepu2018control} as well as the integrity of controller software \cite{patt2019} in the context of industrial control systems. However, these have yet to be generalized to more complex CPS such as drones.


%% file: 8-conclusion.tex
\section{Conclusion}\label{sec:conclusion}
In this paper, we characterized covert data exfiltration over air-gapped cyber-physical channels in the context of edge device applications. In particular, we formalized how an attacker may maintain the stealthiness and utility of a cyber-physical application while maximizing the rate of transmission. We detailed how to practically model attackers and defenders in this context using real-world examples of an industrial control system as well as an autonomous drone surveilling an area. We finally discuss the limitation of current defensive measures and discuss appropriate countermeasures.

%% file: main.bbl
\begin{thebibliography}{10}

\bibitem{google}
Cloud iot edge - extending google cloud's ai \& ml | iot edge | google cloud.

\bibitem{watsonIoT}
Explore the internet of things (iot).

\bibitem{gepredix}
Predix edge computing | edge computing platform | ge digital.

\bibitem{adepu2018control}
{\sc Adepu, S., Brasser, F., Garcia, L., Rodler, M., Davi, L., Sadeghi, A.-R.,
  and Zonouz, S.}
\newblock Control behavior integrity for distributed cyber-physical systems.
\newblock {\em arXiv preprint arXiv:1812.08310\/} (2018).

\bibitem{bertsekas1995dynamic}
{\sc Bertsekas, D.~P., Bertsekas, D.~P., Bertsekas, D.~P., and Bertsekas,
  D.~P.}
\newblock {\em Dynamic programming and optimal control}, vol.~1.
\newblock Athena scientific Belmont, MA, 1995.

\bibitem{bonomi2012fog}
{\sc Bonomi, F., Milito, R., Zhu, J., and Addepalli, S.}
\newblock Fog computing and its role in the internet of things.
\newblock In {\em Proceedings of the first edition of the MCC workshop on
  Mobile cloud computing\/} (2012), ACM, pp.~13--16.

\bibitem{esmaeilzadeh2012neural}
{\sc Esmaeilzadeh, H., Sampson, A., Ceze, L., and Burger, D.}
\newblock Neural acceleration for general-purpose approximate programs.
\newblock In {\em Proceedings of the 2012 45th Annual IEEE/ACM International
  Symposium on Microarchitecture\/} (2012), IEEE Computer Society,
  pp.~449--460.

\bibitem{falliere2011w32}
{\sc Falliere, N., Murchu, L.~O., and Chien, E.}
\newblock W32. stuxnet dossier.
\newblock {\em White paper, Symantec Corp., Security Response 5}, 6 (2011), 29.

\bibitem{faruque2016acoustic}
{\sc Faruque, A., Abdullah, M., Chhetri, S.~R., Canedo, A., and Wan, J.}
\newblock Acoustic side-channel attacks on additive manufacturing systems.
\newblock In {\em Proceedings of the 7th International Conference on
  Cyber-Physical Systems\/} (2016), IEEE Press, p.~19.

\bibitem{freivald1999change}
{\sc Freivald, M.~P., Richards, M.~S., and Noble, A.~C.}
\newblock Change-detection tool indicating degree and location of change of
  internet documents by comparison of cyclic-redundancy-check (crc) signatures,
  Apr.~27 1999.
\newblock US Patent 5,898,836.

\bibitem{garcia2017hey}
{\sc Garcia, L., Brasser, F., Cintuglu, M.~H., Sadeghi, A.-R., Mohammed, O.~A.,
  and Zonouz, S.~A.}
\newblock Hey, my malware knows physics! attacking plcs with physical model
  aware rootkit.
\newblock In {\em NDSS\/} (2017).

\bibitem{garcia2014covert}
{\sc Garcia, L., Senyondo, H., McLaughlin, S., and Zonouz, S.}
\newblock Covert channel communication through physical interdependencies in
  cyber-physical infrastructures.
\newblock In {\em Smart Grid Communications (SmartGridComm), 2014 IEEE
  International Conference on\/} (2014), IEEE, pp.~952--957.

\bibitem{patt2019}
{\sc Ghaeini, H.~R., Chan, M., Bahmani, R., Brasser, F., Garcia, L., Zhou, J.,
  Sadeghi, A.-R., and Zonouz, S.}
\newblock Patt: Physics-based attestation of control systems.
\newblock In {\em International Symposium on Research in Attacks, Intrusions,
  and Defenses\/} (2019), Springer.

\bibitem{giernacki2017crazyflie}
{\sc Giernacki, W., Skwierczy{\'n}ski, M., Witwicki, W., Wro{\'n}ski, P., and
  Kozierski, P.}
\newblock Crazyflie 2.0 quadrotor as a platform for research and education in
  robotics and control engineering.
\newblock In {\em 2017 22nd International Conference on Methods and Models in
  Automation and Robotics (MMAR)\/} (2017), IEEE, pp.~37--42.

\bibitem{gremban2019}
{\sc Gremban, K.}
\newblock What is azure iot edge.

\bibitem{gur2018xled}
{\sc Gur, M., Zadov, B., Daidakulov, A., and Elovici, Y.}
\newblock xled: Covert data exfiltration from air-gapped networks via switch
  and router leds.
\newblock In {\em 2018 16th Annual Conference on Privacy, Security and Trust
  (PST)\/} (2018), IEEE, pp.~1--12.

\bibitem{guri2019air}
{\sc Guri, M., and Bykhovsky, D.}
\newblock air-jumper: Covert air-gap exfiltration/infiltration via security
  cameras \& infrared (ir).
\newblock {\em Computers \& Security 82\/} (2019), 15--29.

\bibitem{guri2018magneto}
{\sc Guri, M., Daidakulov, A., and Elovici, Y.}
\newblock Magneto: Covert channel between air-gapped systems and nearby
  smartphones via cpu-generated magnetic fields.
\newblock {\em arXiv preprint arXiv:1802.02317\/} (2018).

\bibitem{guri2015gsmem}
{\sc Guri, M., Kachlon, A., Hasson, O., Kedma, G., Mirsky, Y., and Elovici, Y.}
\newblock Gsmem: Data exfiltration from air-gapped computers over gsm
  frequencies.
\newblock In {\em USENIX Security Symposium\/} (2015), pp.~849--864.

\bibitem{guri2014airhopper}
{\sc Guri, M., Kedma, G., Kachlon, A., and Elovici, Y.}
\newblock Airhopper: Bridging the air-gap between isolated networks and mobile
  phones using radio frequencies.
\newblock In {\em Malicious and Unwanted Software: The Americas (MALWARE), 2014
  9th International Conference on\/} (2014), IEEE, pp.~58--67.

\bibitem{guri2016usbee}
{\sc Guri, M., Monitz, M., and Elovici, Y.}
\newblock Usbee: Air-gap covert-channel via electromagnetic emission from usb.
\newblock In {\em Privacy, Security and Trust (PST), 2016 14th Annual
  Conference on\/} (2016), IEEE, pp.~264--268.

\bibitem{guri2015bitwhisper}
{\sc Guri, M., Monitz, M., Mirski, Y., and Elovici, Y.}
\newblock Bitwhisper: Covert signaling channel between air-gapped computers
  using thermal manipulations.
\newblock In {\em Computer Security Foundations Symposium (CSF), 2015 IEEE
  28th\/} (2015), IEEE, pp.~276--289.

\bibitem{guri2016fansmitter}
{\sc Guri, M., Solewicz, Y., Daidakulov, A., and Elovici, Y.}
\newblock Fansmitter: Acoustic data exfiltration from (speakerless) air-gapped
  computers.
\newblock {\em arXiv preprint arXiv:1606.05915\/} (2016).

\bibitem{guri2017acoustic}
{\sc Guri, M., Solewicz, Y., Daidakulov, A., and Elovici, Y.}
\newblock Acoustic data exfiltration from speakerless air-gapped computers via
  covert hard-drive noise (‘diskfiltration’).
\newblock In {\em European Symposium on Research in Computer Security\/}
  (2017), Springer, pp.~98--115.

\bibitem{guri2018mosquito}
{\sc Guri, M., Solwicz, Y., Daidakulov, A., and Elovici, Y.}
\newblock Mosquito: Covert ultrasonic transmissions between two air-gapped
  computers using speaker-to-speaker communication.
\newblock {\em arXiv preprint arXiv:1803.03422\/} (2018).

\bibitem{guri2018powerhammer}
{\sc Guri, M., Zadov, B., Bykhovsky, D., and Elovici, Y.}
\newblock Powerhammer: Exfiltrating data from air-gapped computers through
  power lines.
\newblock {\em arXiv preprint arXiv:1804.04014\/} (2018).

\bibitem{guri2018odini}
{\sc Guri, M., Zadov, B., Daidakulov, A., and Elovici, Y.}
\newblock Odini: Escaping sensitive data from faraday-caged, air-gapped
  computers via magnetic fields.
\newblock {\em arXiv preprint arXiv:1802.02700\/} (2018).

\bibitem{guri2017led}
{\sc Guri, M., Zadov, B., and Elovici, Y.}
\newblock Led-it-go: Leaking (a lot of) data from air-gapped computers via the
  (small) hard drive led.
\newblock In {\em International Conference on Detection of Intrusions and
  Malware, and Vulnerability Assessment\/} (2017), Springer, pp.~161--184.

\bibitem{gustafson1975design}
{\sc Gustafson, D.~E., and Speyer, J.~L.}
\newblock Design of linear regulators for nonlinear stochastic systems.
\newblock {\em Journal of Spacecraft and Rockets 12}, 6 (1975), 351--358.

\bibitem{ha2014towards}
{\sc Ha, K., Chen, Z., Hu, W., Richter, W., Pillai, P., and Satyanarayanan, M.}
\newblock Towards wearable cognitive assistance.
\newblock In {\em Proceedings of the 12th annual international conference on
  Mobile systems, applications, and services\/} (2014), ACM, pp.~68--81.

\bibitem{kopf2007informationf}
{\sc K{\"o}pf, B., and Basin, D.}
\newblock An information-theoretic model for adaptive side-channel attacks.
\newblock In {\em Proceedings of the 14th ACM conference on Computer and
  communications security\/} (2007), ACM, pp.~286--296.

\bibitem{kurniawan2018learning}
{\sc Kurniawan, A.}
\newblock {\em Learning AWS IoT: Effectively manage connected devices on the
  AWS cloud using services such as AWS Greengrass, AWS button, predictive
  analytics and machine learning}.
\newblock Packt Publishing Ltd, 2018.

\bibitem{papernot2016limitations}
{\sc Papernot, N., McDaniel, P., Jha, S., Fredrikson, M., Celik, Z.~B., and
  Swami, A.}
\newblock The limitations of deep learning in adversarial settings.
\newblock In {\em 2016 IEEE European Symposium on Security and Privacy
  (EuroS\&P)\/} (2016), IEEE, pp.~372--387.

\bibitem{shi2016edge}
{\sc Shi, W., Cao, J., Zhang, Q., Li, Y., and Xu, L.}
\newblock Edge computing: Vision and challenges.
\newblock {\em IEEE Internet of Things Journal 3}, 5 (2016), 637--646.

\bibitem{song2016my}
{\sc Song, C., Lin, F., Ba, Z., Ren, K., Zhou, C., and Xu, W.}
\newblock My smartphone knows what you print: Exploring smartphone-based
  side-channel attacks against 3d printers.
\newblock In {\em Proceedings of the 2016 ACM SIGSAC Conference on Computer and
  Communications Security\/} (2016), ACM, pp.~895--907.

\bibitem{srinivasan2016privacy}
{\sc Srinivasan, R., Mohan, A., and Srinivasan, P.}
\newblock Privacy conscious architecture for improving emergency response in
  smart cities.
\newblock In {\em 2016 Smart City Security and Privacy Workshop (SCSP-W)\/}
  (2016), IEEE, pp.~1--5.

\bibitem{sun2019mismo}
{\sc Sun, P., Garcia, L., and Zonouz, S.}
\newblock Tell me more than just assembly! reversing cyber-physical execution
  semantics of embedded iot controller software binaries.
\newblock In {\em 2019 49th Annual IEEE/IFIP International Conference on
  Dependable Systems and Networks (DSN)\/} (2019), IEEE.

\bibitem{yao2018fastdeepiot}
{\sc Yao, S., Zhao, Y., Shao, H., Liu, S., Liu, D., Su, L., and Abdelzaher, T.}
\newblock Fastdeepiot: Towards understanding and optimizing neural network
  execution time on mobile and embedded devices.
\newblock In {\em Proceedings of the 16th ACM Conference on Embedded Networked
  Sensor Systems\/} (2018), ACM, pp.~278--291.

\bibitem{yi2015fog}
{\sc Yi, S., Hao, Z., Qin, Z., and Li, Q.}
\newblock Fog computing: Platform and applications.
\newblock In {\em 2015 Third IEEE Workshop on Hot Topics in Web Systems and
  Technologies (HotWeb)\/} (2015), IEEE, pp.~73--78.

\end{thebibliography}
